\documentclass[a4paper,11pt]{amsart}
\usepackage{etoolbox}

\makeatletter
\patchcmd{\@setauthors}{\MakeUppercase}{}{}{}
\makeatother

\usepackage{multicol}
\usepackage{multirow}
\usepackage{amsmath,latexsym,amsbsy,amssymb,fancyhdr}
\usepackage{array}
\usepackage{enumerate}
\usepackage{amsthm}
\usepackage{booktabs} 
\usepackage{multirow}
\usepackage{tabularx}
 \usepackage{epsfig} 

\usepackage{psfrag,graphicx}
\usepackage{epstopdf}
\usepackage[latin1]{inputenc}
\usepackage[colorlinks=true]{hyperref}
\hypersetup{urlcolor=blue, citecolor=blue, linkcolor=blue}\usepackage{multicol}
\usepackage{multirow}
\usepackage{amsmath,latexsym,amsbsy,amssymb,fancyhdr}
\usepackage{caption}
\usepackage{subcaption}

\usepackage{array}
\usepackage{enumerate}
\usepackage{amsthm}
\usepackage{booktabs} 
\usepackage{multirow}
\usepackage{tabularx}
 \usepackage{epsfig} 

\usepackage{psfrag,graphicx}
\usepackage{epstopdf}
\usepackage[latin1]{inputenc}
\usepackage[colorlinks=true]{hyperref}
\hypersetup{urlcolor=blue, citecolor=blue, linkcolor=blue}

\usepackage{color}
\usepackage{cite}
\usepackage{commath}
\usepackage{mathtools}
\usepackage{systeme}

\newcommand{\at}[2][]{#1|_{#2}}

\numberwithin{equation}{section}

\makeatletter
 
\let\insertdate\@date
\makeatother

\usepackage{xcolor}
\definecolor{palegreen}{rgb}{0.2,0.6,0.2}

\author{Binh Duc Truong, Cuong Phu Le and Einar Halvorsen}
\address{Department of Microsystems, University College of Southeast Norway,
Campus Vestfold, Raveien 215, N-3184 Borre, Norway. Email: Binh.Truong@usn.no}
\begin{document}
\title{Theoretical analysis of electrostatic energy harvester configured as Bennet's doubler based on Q-V cycles}
\maketitle


\keywords{\textbf{\textit{Abstract --}} This paper presents theoretical analysis of a MEMS electrostatic energy harvester configured as the Bennet's doubler. Steady-state operation of the doubler circuit can be approximated by a right-angled trapezoid Q-V cycle. A similarity between voltage doubler and resistive-based charge-pump circuit is highlighted. By taking electromechanical coupling into account, the analytical solution of the saturation voltage is the first time derived, providing a greater comprehension of the system performance and multi-parameter effects. The theoretical approach is verified by results of circuit simulation for two cases of mathematically idealized diode and of Schottky diode. Development of the doubler/multiplier circuits that can further increase the saturation voltage is investigated.}

\section{Introduction}
Wireless sensor nodes (WSNs) are emerging as one of the most commonly used monitoring and sensing systems \cite{Sergiou2014, Khan2016}. Currently, most WSNs are powered by batteries. Energy harvesting from vibration becomes a potential alternative to obtain electrical energy for WSNs, especially in some circumstances where batteries may not be feasible. For the vibration energy harvesters, there are three common transduction mechanisms which includes piezoelectric, electromagnetic and electrostatic \cite{duToit2005, Beeby2007, Chiu2007}. In this paper, we focus on the electrostatic energy harvesting system.

One of the problems associated with the electrostatic energy harvesters is the implementation of power management circuits. As an example, a conversion circuit consisted of a voltage source, a variable capacitor and two switches was presented in \cite{Roundy2003, Dragunov2012}. 
Although energy transduction through this circuit is possible, the regime where the output voltage saturated was not discussed. Several solutions based on energy-renewal technique for extracting electrical energy were presented. 
For instance, Yen \textit{et al.} proposed a configuration of single variable-capacitance harvester, combining an asynchronous charge-pump with an inductive fly-back circuit to recharge the scavenging capacitor \cite{Yen2006}. Mitcheson \textit{et al.} developed a buck-boost topology with bi-directional switches for rectifying and increasing the AC voltage obtained from a transducer \cite{Mitcheson2007}. These circuit topologies face the trade-off between power consumption of control unit and harvester efficiency.

The Bennet's doubler was early introduced in 1787 by the Reverend Bennet and Kaye \cite{Bennet1787}. The device is used for the continuous doubling of an initial small charge through a sequence of operations with three plates. 
Based on this approach, de Queiroz proposed a promising variation of such a voltage doubler for macro-scale vibration energy harvesters composed by variable capacitors and diodes \cite{deQueiroz2011, deQueiroz2011b, deQueiroz2013}. In order to adapt the concept to micro-scale electrostatic generators, several researches have been developed and investigated \cite{Dragunov2013, Lefeuvre2014, Dorzhiev2015, Lefeuvre2015June}, including attempts to increase the charging current for a reservoir capacitor or to optimize the harvested power.
In a recent work by Galayko \cite{Galayko2016}, operation of the doubler configuration with a single variable capacitor was thoroughly analyzed in the electrical domain. The shape of Q-V diagram obtained from simulation is very close to be rectangular.
However, operation of a transducer configuration with two time-varying capacitors and the dependence of the saturated voltage on dynamic characteristics of the mechanical domain has not explored yet.

Since the saturation phenomenon was observed in experiments \cite{Dorzhiev2015}, the effect of the electromechanical coupling on it is of interest to study. This paper further presents a theoretical analysis of the Bennet's doubler based on the Q-V cycle. A complete model of an anti-phase overlap-varying transducers electrically configured as a voltage doubler is investigated. Numerical results for both ideal- and non-ideal diodes are obtained by means of a SPICE simulator, which are used to support the analytical solutions. For further increase of the saturated voltage across the storage capacitor, alternative topologies are introduced and analyzed.

\section{Steady state operation with mathematically idealized diodes} \label{Ideal_Diode}

\subsection{Theoretical analysis}

\begin{figure}[!htbp]
	\centering
	\includegraphics[width=0.45\textwidth]{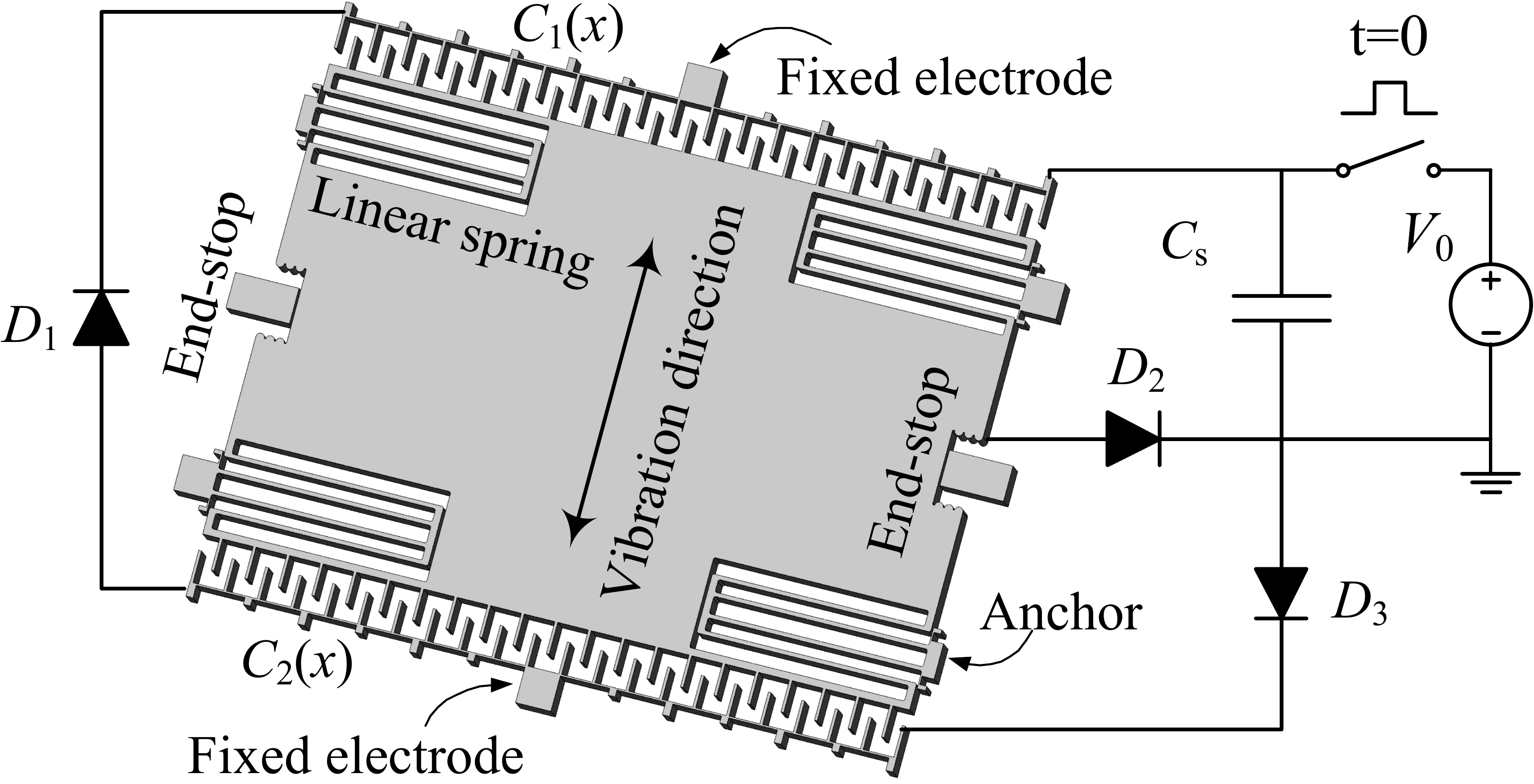}
	\caption{Overlap-varying energy harvesters employing the Bennet's doubler circuit.}
	\label{Fig:Device}
\end{figure}
The overlap-varying energy harvesters can be utilized in a charge-doubling circuit-configuration as shown in Figure \ref{Fig:Device}. The proof mass is suspended by four folded-beam linear springs. The maximum displacement $X_\mathrm{max}$ is defined by the mechanical end-stops.
Two anti-phase variable capacitors $C_{1/2} (x) = C_0 (1  \mp \frac{x}{x_0})$ are connected to three diodes $D_1,\, D_2, \, D_3$ and the storage capacitor $C_\mathrm{s}$. Here $C_0,\, x_0$ and $x$ are the nominal capacitance, the nominal overlap and the proof mass displacement respectively. Operation of the doubler circuit does not require any control unit or switches but an initial bias voltage $V_0$.

\begin{figure}[!htbp]
	\centering
	\includegraphics[width=0.4\textwidth]{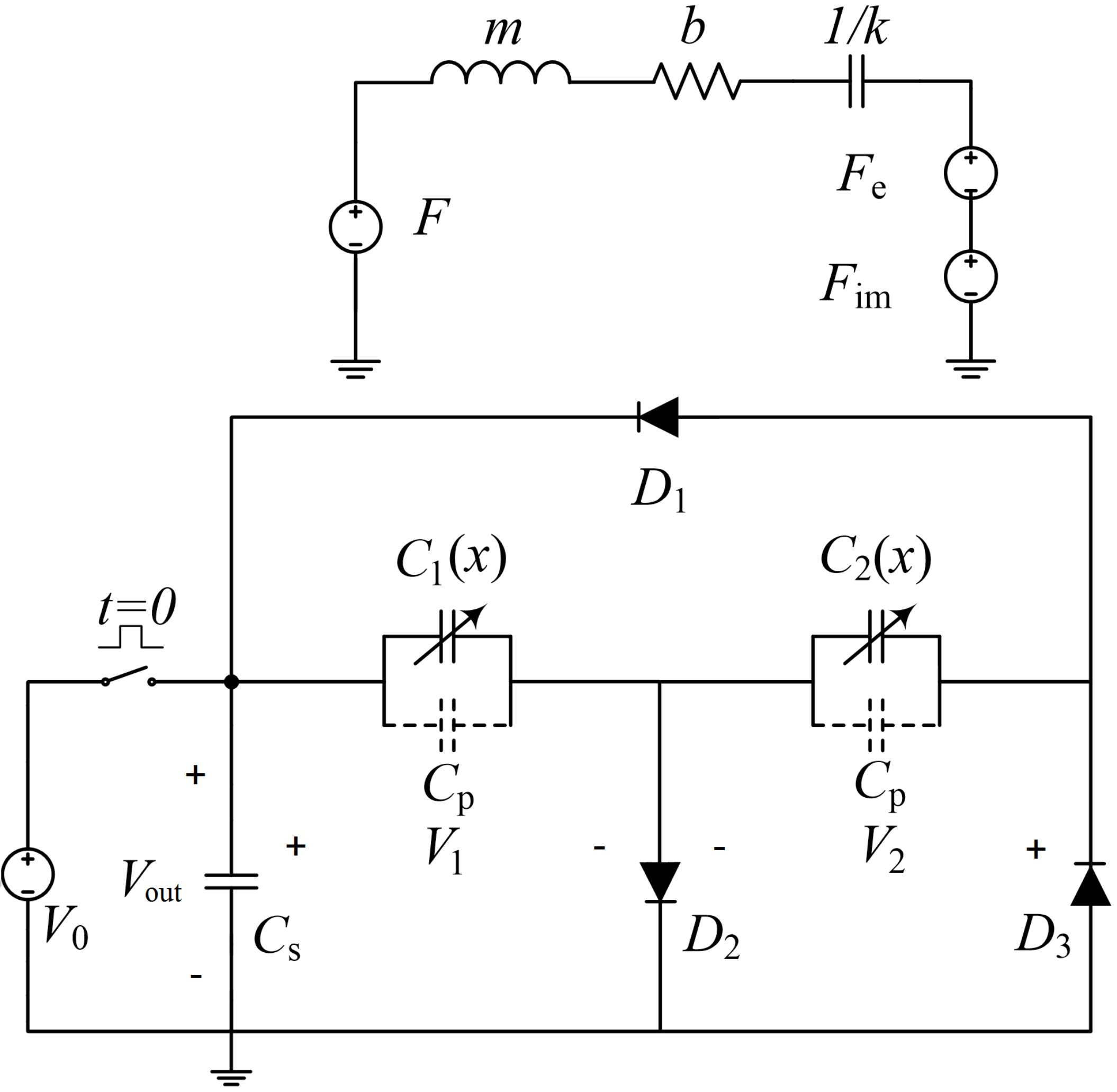}
	\caption{Equivalent circuit for mechanical domain and Bennet's doubler configuration.}
	\label{Fig:Lumped-model}
\end{figure}
Figure \ref{Fig:Lumped-model} shows a complete lumped-model of the doubler configuration including equivalent circuit for the mechanical subsystem, where $m$ - proof mass, $b$ - mechanical damping, $k$ - total spring stiffness, $F$ - an external force, $F_\mathrm{e}$ - the electrostatic force and $C_\mathrm{p}$ - the parasitic capacitance of each transducer. The contact force $F_\mathrm{im}$ is simply modeled as a spring-damper system $F_\mathrm{im} = k_\mathrm{im}  \delta  + b_\mathrm{im} \dot{\delta} $ for $\abs{x} \geq X_\mathrm{max}$ \cite{Le2012}, where $\delta =  \abs{x}-X_\mathrm{max}$ is relative displacement between the proof mass and the end-stops, $k_\mathrm{im}$ is the impact stiffness and $b_\mathrm{im}$ is the impact damping.
\begin{figure}[!thbp]
	\centering
	\includegraphics[width=0.45\textwidth]{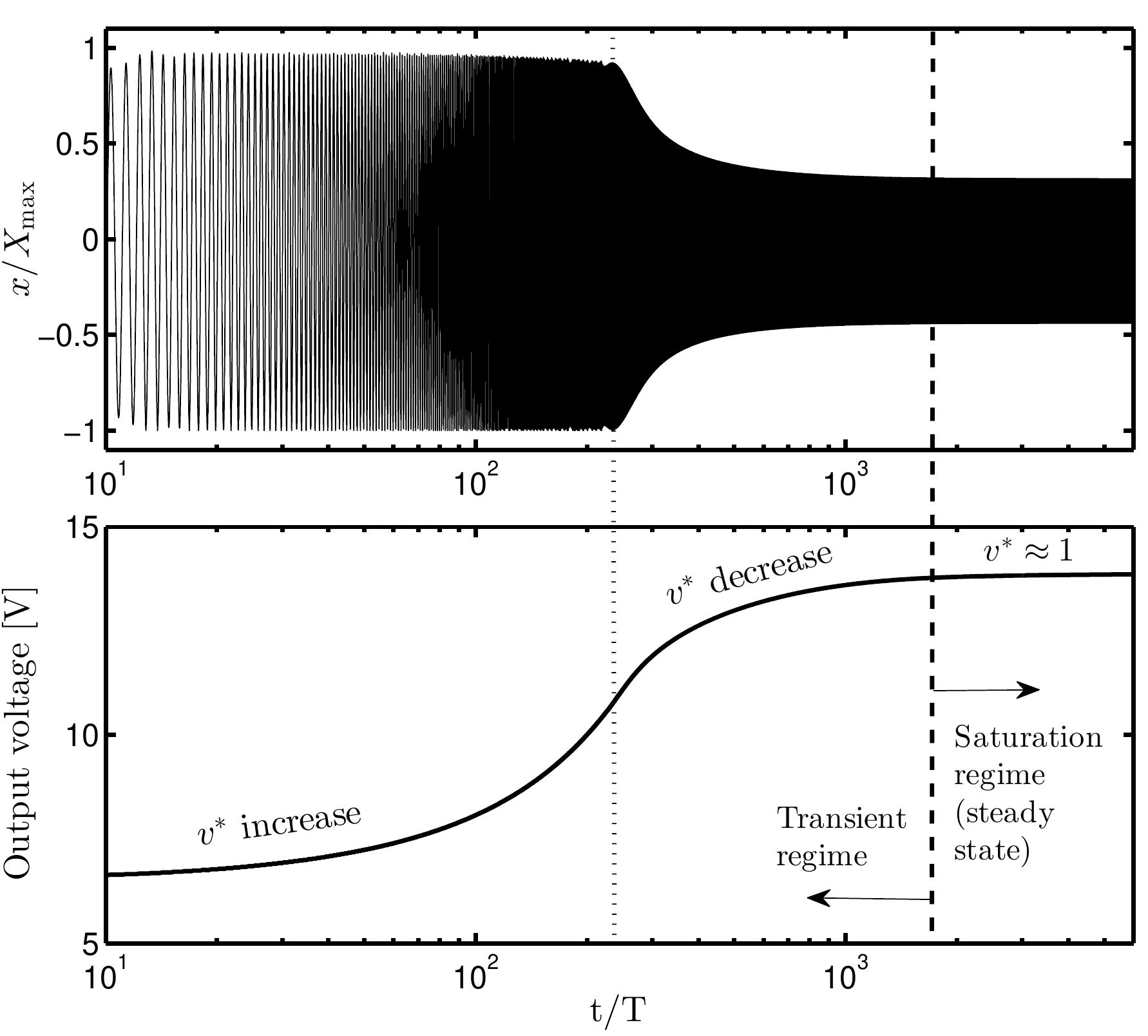}
	\caption{Evolution of the proof mass displacement and the output voltage across the storage capacitor with the input acceleration amplitude $A=2.0$ g, the drive frequency $f=f_0$ and the initial bias voltage $V_0 = 7$ V.}
	\label{Fig:Vs_Disp_Full}
\end{figure}
For a sufficient voltage $V_0$ and an adequate input acceleration amplitude $A$, the voltage accumulated on the storage capacitor $C_\mathrm{s}$ initially increases. The vibration frequency is chosen $f=f_0=\frac{1}{2 \pi} \sqrt{\frac{k}{m}}$. 
Figure \ref{Fig:Vs_Disp_Full} shows that after certain cycles of transient regime, the steady state is achieved. The electrical energy is no longer harvested and the output voltage $V_\mathrm{out}$ is then maintained constant at $V_\mathrm{s}$ (i.e., saturation voltage).

\begin{table}
	\centering
	\Small
	\caption{Model parameters}%
	\begin{tabular}{l l l} 
		\hline\hline
		\textbf{Parameters} & \textbf{Value} \\
		\hline
		Proof mass, $m$ & 1.022 mg\\ 
		Spring stiffness, $k$ & 3.595 N/m\\ 
		Thin-film air damping, $b$ & 3.478e-5 Ns/m \\
		Nominal overlap, $x_0$ & 80 $\mu$m \\
		Nominal capacitance, $C_\mathrm{0}$ & 15 pF \\
		Parasitic capacitance, $C_\mathrm{p}$ & 7.5 pF \\
		Storage capacitance, $C_\mathrm{s}$ & 10 nF \\
		Contact stiffness, $k_\mathrm{im}$ & 3.361 MN/m \\
		Impact damping, $b_\mathrm{im}$ & 0.435 Ns/m \\
		Maximum displacement, $X_\mathrm{max}$ & 80 $\mu$m \\
		\hline\hline
	\end{tabular}
	\label{Table:Model_Params} 
\end{table}
The proof mass displacement amplitude $X_0$ changes in complicated manner: $X_0$ first reaches the maximum value $X_0 \approx X_\mathrm{max}$ (i.e., which is limited by the mechanical end-stops), then decreases and kept fixed at $X_0 \approx X_\mathrm{s}$ in saturation regime.
For convenience, we define the rate of voltage evolution $v^\ast$ as a ratio of the maximum output voltage in two subsequent period
\begin{align}
\small
v^\ast = \frac{\mathrm{max} \Big( V_\mathrm{out} \at[\Big]{T_\mathrm{i+1} }\Big) }{\mathrm{max} \Big( V_\mathrm{out} \at[\Big]{T_\mathrm{i} }\Big)}.
\end{align}
As shown in Figure \ref{Fig:Vs_Disp_Full}, $v^\ast$ is modified over cycles under the variation of $X_0$ as follows. $v^\ast$ is small at the beginning and gradually increases, meanwhile $X_0 \approx X_\mathrm{max}$. After reaching the maximum, $v^\ast$ decreases with reduction of $X_0$ and finally becomes one at steady state.
Ultimately higher voltages through the conversion phase induce more effective electrical damping represented by electrostatic force in the transducers, causing a decrease of the proof mass displacement. As a consequence, the transducer capacitance ratio is reduced to $\eta = (C_\mathrm{max}+C_\mathrm{p})/(C_\mathrm{min}+C_\mathrm{p}) \approx 1.72$, which is no more satisfied the condition of the doubler circuit operation $\eta_\mathrm{cr} = 2$. Therefore, $V_\mathrm{out}$ is saturated at a certain value.
Detail of dynamic analyses and the model parameters (i.e., listed in Table \ref{Table:Model_Params}) are referred to \cite{TruongSS17}.
In this paper, the effect of the electrostatic force on $V_\mathrm{s}$ is the major objective of investigation.

\begin{figure}[!thbp]
	\centering
	\includegraphics[width=0.45\textwidth]{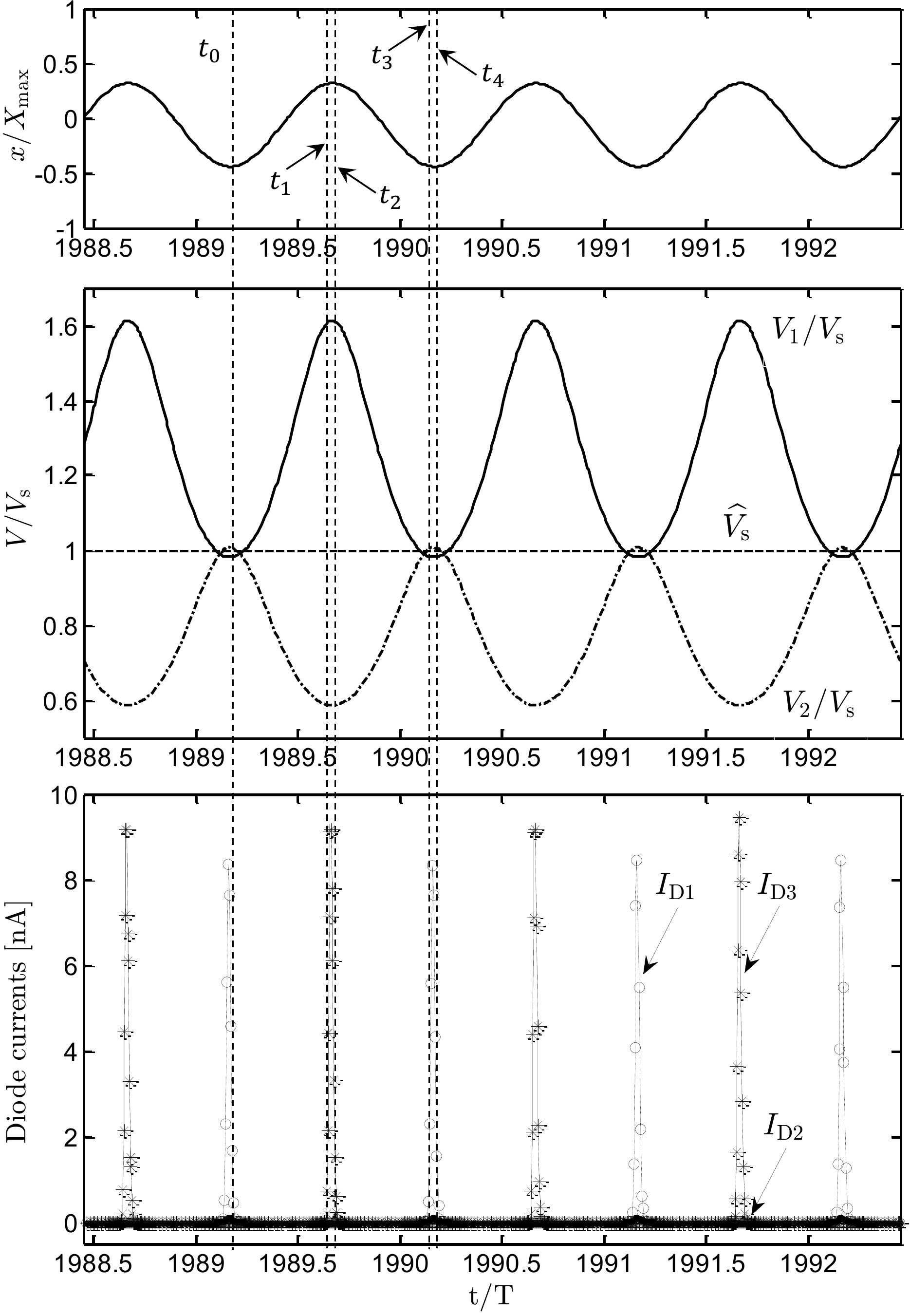}
	\caption{Waveforms of displacement, voltages on variable capacitors and currents through three diodes at steady state with $A=2.0$ g and $f=f_0$.}
	\label{Fig:Math_wave}
\end{figure}
\begin{figure}[!htbp]
	\centering
	\includegraphics[width=0.35\textwidth]{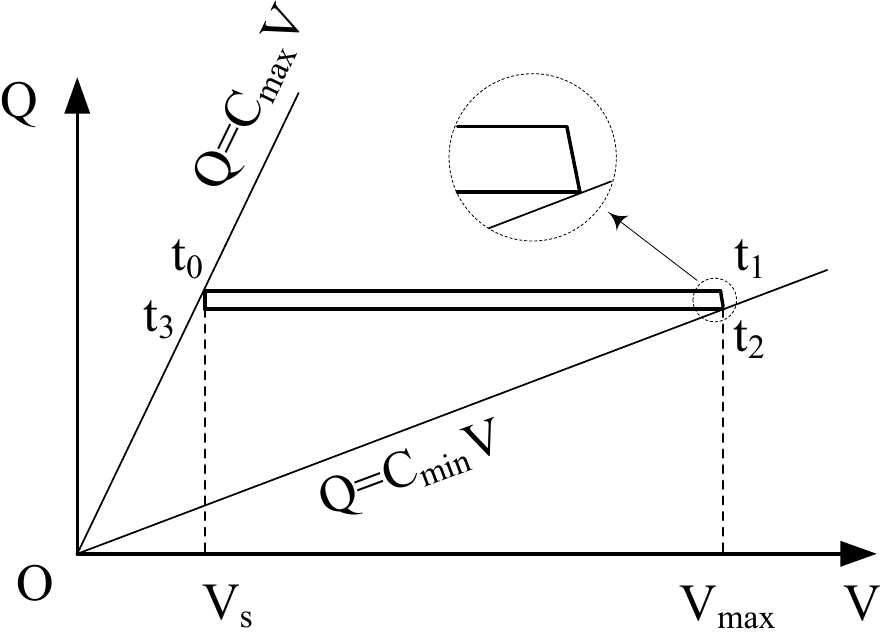}
	\caption{Approximated Q-V diagram of variable capacitor $C_1 (x)$ at steady state with mathematically ideal diodes.}
	\label{Fig:QV_ideal}
\end{figure}
Figure \ref{Fig:Math_wave} shows waveforms of the proof mass displacement, the voltages $V_1, \,V_2$ across $C_\mathrm{1}, \,C_\mathrm{2}$ and the currents $I_\mathrm{D1}, \, I_\mathrm{D2}, \, I_\mathrm{D3}$ through three mathematically idealized diodes respectively. Operation of the doubler circuit at steady state can be divided into a sequence of four stages from $t_0$ to $t_4$.
Based on the dynamic simulations, we observe that the relation of $Q_1$ and $V_1$ at steady state can be approximated by a right-angled trapezoid Q-V cycle diagram and the time interval between $\Delta t_{21}=t_2 - t_1$ and $\Delta t_{43}=t_4 - t_3$ are very small, as depicted in Figure \ref{Fig:QV_ideal}.

\textbf{\textit{Stage I:}}

At $t=t_0$, $x (t_0) = - X_\mathrm{s}$ and $V_1 (t_0) \approx V_2 (t_0) \approx V_\mathrm{s}$, where $X_\mathrm{s}$ is the maximum displacement at steady state. 
From $t_0$ to $t_1$, all three diodes $D_1, \, D_2$ and $D_3$ are blocking as the condition $V_\mathrm{C_\mathrm{2}} < V_\mathrm{0} < V_\mathrm{C_\mathrm{1}} < V_\mathrm{C_\mathrm{2}} + V_\mathrm{0}$ is satisfied.
The charges on the two transducers are
\begin{align}
\small
q_1 (t_0) &= V_\mathrm{s} \Big[C_\mathrm{p} + C_0 (1 + \frac{X_\mathrm{s}}{x_0})\Big] \label{q1_t0}, \\
q_2 (t_0) &= V_\mathrm{s} \Big[C_\mathrm{p} + C_0 (1 - \frac{X_\mathrm{s}}{x_0})\Big] \label{q2_t0}.
\end{align}

In the first stage, $q_1$ and $q_2$ are constants, $V_1$ and $V_2$ are given
\begin{align}
\small
V_1 \at[\Big]{t \in [t_0,\, t_1]} &= \frac{q_1}{C_1} = \frac{ V_\mathrm{s} \Big[C_\mathrm{p} + C_0 (1 + \frac{X_\mathrm{s}}{x_0})\Big]}{C_\mathrm{p} + C_0 (1 - \frac{x}{x_0})} \label{V_1} , \\
V_2 \at[\Big]{t \in [t_0,\, t_1]}&= \frac{q_2}{C_2} = \frac{ V_\mathrm{s} \Big[C_\mathrm{p} + C_0 (1 - \frac{X_\mathrm{s}}{x_0})\Big]}{C_\mathrm{p} + C_0 (1 + \frac{x}{x_0})} \label{V_2} .
\end{align}

\textbf{\textit{Stage II:}}

At $t=t_1$, $V_1 (t_1) \approx V_2 (t_1) + V_\mathrm{s}$ and diode $D_3$ starts to conduct. Since the time interval between $t_1$ and $t_2$ is very small (i.e., see Figure \ref{Fig:QV_ideal}), the proof mass displacement at $t_1$ can be approximated as $x(t_1) \approx x(t_2) = X_\mathrm{s}$, then
\begin{align}
\small
\frac{1 + \frac{C_0}{C_\mathrm{p}}\Big( 1 + \frac{X_\mathrm{s}}{x_0} \Big)}{1 + \frac{C_0}{C_\mathrm{p}} \Big( 1 - \frac{X_\mathrm{s}}{x_0} \Big)} = 1 + \frac{1 + \frac{C_0}{C_\mathrm{p}}\Big( 1 - \frac{X_\mathrm{s}}{x_0} \Big)}{1 + \frac{C_0}{C_\mathrm{p}} \Big( 1 + \frac{X_\mathrm{s}}{x_0} \Big)}.
\label{D3_ideal}
\end{align}
The solution is given as
\begin{align}
\small
X_\mathrm{s} = 3 \big(\frac{\sqrt{5}}{2} - 1\big) x_0 .
\label{Xs_ideal}
\end{align}
The peak values of voltages across $C_1$ and $C_2$ are
\begin{align}
\small
V_\mathrm{I} &= V_1 (t_1) = V_1 (t_2) =  V_\mathrm{s} \frac{\sqrt{5} + 1}{2},
\label{V1_peak_ideal} \\
V_\mathrm{II} &= V_2 (t_1) = V_2 (t_2)= V_\mathrm{s} \frac{\sqrt{5} - 1}{2}. \label{V2_peak_ideal}
\end{align}

In this stage, charges $\Delta Q_\mathrm{s}$ and $\Delta Q$ are pumped from $C_1$ into $C_\mathrm{s}$ and $C_2$ respectively. At steady state, $V_\mathrm{s}$ is considered unchanged, thus $\Delta Q_\mathrm{s}$ is neglected.

\textbf{\textit{Stage III:}}

From $t_2$ to $t_3$, all diodes are blocked, $q_1$ and $q_2$ are constants
\begin{align}
q_1 \at[\Big]{t \in [t_2,\, t_3]} &= q_1(t_2) = V_\mathrm{s} \Big[C_\mathrm{p} + C_0 (1 + \frac{X_\mathrm{s}}{x_0})\Big] - \Delta Q , \label{Ideal:q1_t2_t3} \\
q_2 \at[\Big]{t \in [t_2,\, t_3]}&= q_2(t_2) = V_\mathrm{s} \Big[C_\mathrm{p} + C_0 (1 - \frac{X_\mathrm{s}}{x_0})\Big] + \Delta Q .
\label{Ideal:q2_t2_t3}
\end{align}

At $t_3$, $x(t_3) = x_3$, $V_1(t_3) = V_\mathrm{s}~ \refstepcounter{equation}(\theequation) \label{Ideal:V1_Vs}$ and $D_2$ starts to conduct transferring amount of charge $\Delta Q^\ast$ from $C_\mathrm{s}$ into $C_1$. Similarly, since $V_s$ is treated as constant, $\Delta Q^\ast$ is thus negligible. The relation \eqref{Ideal:V1_Vs} now can be written as
\begin{align}
V_1 (t_3) = \frac{q_1(t_3)}{C_1 (t_3)} = \frac{ V_\mathrm{s} \Big[C_\mathrm{p} + C_0 (1 + \frac{X_\mathrm{s}}{x_0})\Big] - \Delta Q}{C_\mathrm{p} + C_0 (1 - \frac{x_3}{x_0})} = V_\mathrm{s} .
\label{Ideal:V1_t3}
\end{align}
Due to the small interval time between $t_3$ and $t_4$, $x_3 \approx x(t_4) = -X_\mathrm{s}$, resulting in $\Delta Q \approx 0$. In other words, the charge transfered from $C_1$ into $C_2$ is insignificant.

Considering the voltage across the capacitor $C_2$ at $t_3$
\begin{align}
\begin{split}
V_2 (t_3) = \frac{q_2(t_3)}{C_2 (t_3)} = \frac{ V_\mathrm{s} \Big[C_\mathrm{p} + C_0 (1 - \frac{X_\mathrm{s}}{x_0})\Big] + \Delta Q}{C_\mathrm{p} + C_0 (1 + \frac{x_3}{x_0})} \approx V_\mathrm{s} .
\end{split}
\end{align}
Therefore, $D_1$ also starts to conduct at $t_3$ since the condition $V_2 \approx V_\mathrm{s}$ holds.

\textbf{\textit{Stage IV:}}

From $t_3$ to $t_4$, $D_1$ is conducting and $\Delta Q$ is transfered from $C_2$ into $C_1$. The charge $q_4$ is
\begin{align}
q_1(t_4) = q_1(t_3) + \Delta Q = V_\mathrm{s} \Big[C_\mathrm{p} + C_0 (1 + \frac{X_\mathrm{s}}{x_0})\Big] .
\label{Ideal:q1_t4}
\end{align}

The condition $q_1 (t_4) = q_1 (t_0) ~ \refstepcounter{equation}(\theequation) \label{Ideal:q1_t4_t0}$ is fulfilled, showing that the state of the doubler circuit at $t_4$ is exactly the same as when $t=t_0$, and a new cycle starts. This also proves that the right-angled trapezoid Q-V cycle diagram is capable of describing the operation of the doubler circuit.

\subsection{Similarity of Bennet'doubler and charge-pump circuit}

Among different circuit topologies for the interface electronics of MEMS capacitive energy harvesters \cite{Burrow2013, Burrow2015}, the charge pump circuit early presented by Roundy \textit{et al.} \cite{Roundy2002} is one of the most promising topologies. Another variation with inductive fly-back circuitry was developed by Yen \textit{et al.} \cite{Yen2006}.
The simplest way to implement fly-back is to use a load resistance, originally reported in \cite{Florentino2011}. Such a fly-back configuration was thoroughly analyzed in \cite{O'Riordan2015}. 

Comparing the results shown in the literature with the one obtained in this paper, it is worth to note that the Q-V cycle for the charge pump circuit with resistive fly-back is very similar to that of Bennet's doubler circuit. Both topologies can be approximated by trapezoidal conversion cycle. At the steady state of the idealized charge pump and the voltage doubler, the Q-V cycle is degenerated to a line (i.e., see Figure \ref{Fig:QV_ideal}).

\section{Approximation of the saturation voltage with mathematically ideal diode}

The electrostatic force $F_\mathrm{e}$ plays an important role in saturation of the output voltage and is thoroughly analyzed in this Section. $F_\mathrm{e}$ is modeled as
\begin{align}
F_\mathrm{e} = -\frac{\partial W_\mathrm{e}}{\partial x} = - \frac{1}{2} \frac{\partial C_1 (x)}{\partial x} V_1^2 - \frac{1}{2} \frac{\partial C_2 (x)}{\partial x} V_2^2 = \frac{1}{2} \frac{C_0}{x_0} \big(V_1^2 - V_2^2\big) 
\end{align}
where $W_\mathrm{e}$ is the electrostatic energy of the transducers. $V_1$ and $V_2$ can be simplified as anti-phase sinusoidal signals for the sake of analysis although it is more complicated than that in reality. Based on the dynamic simulations, we observe that the phase difference between \textit{the input acceleration} and \textit{the voltage across $C_1$} is negligibly small and is ignored.
The waveforms of $V_1$ and $V_2$ are then presented as
\begin{align}
V_1 &= \frac{V_\mathrm{I} + V_\mathrm{s}}{2} + \frac{V_\mathrm{I} - V_\mathrm{s}}{2} \sin (\omega t) = V_\mathrm{s} \frac{3 + \sqrt{5}}{4} + V_\mathrm{s} \frac{-1 + \sqrt{5}}{4} \sin (\omega t) ,
\label{V1_ideal} \\
V_2 &= \frac{V_\mathrm{II} + V_\mathrm{s}}{2} - \frac{V_\mathrm{s} - V_\mathrm{II}}{2} \sin (\omega t) = V_\mathrm{s} \frac{1 + \sqrt{5}}{4} - V_\mathrm{s} \frac{3 - \sqrt{5}}{4} \sin (\omega t)
\label{V2_ideal}
\end{align}
yielding
\begin{align}
V_1^2 - V_2^2 & = \frac{2 + \sqrt{5}}{4} V_\mathrm{s}^2 \big( 1 + \sin (\omega t) \big) \Big ( 1 + \frac{(\sqrt{5}-2)^2}{2} \sin (\omega t) \Big) .
\end{align}
The coefficient $\frac{(\sqrt{5}-2)^2}{2} \approx 0.028 \ll 1$ is negligible, the electrostatic force is then
\begin{align}
\small
F_\mathrm{e} = \frac{2 + \sqrt{5}}{8} \frac{C_0}{x_0} V_\mathrm{s}^2 \big( 1 + \sin (\omega t)\big) = F_0 \big( 1 + \sin (\omega t)\big) .
\label{Eq:Fe}
\end{align}
where $F_0 = \frac{2 + \sqrt{5}}{8} \frac{C_0}{x_0} V_\mathrm{s}^2$. The harmonic term of $F_\mathrm{e}$ is in phase with the input acceleration.
\begin{figure}[!tbp]
	\centering
	\includegraphics[width=0.45\textwidth]{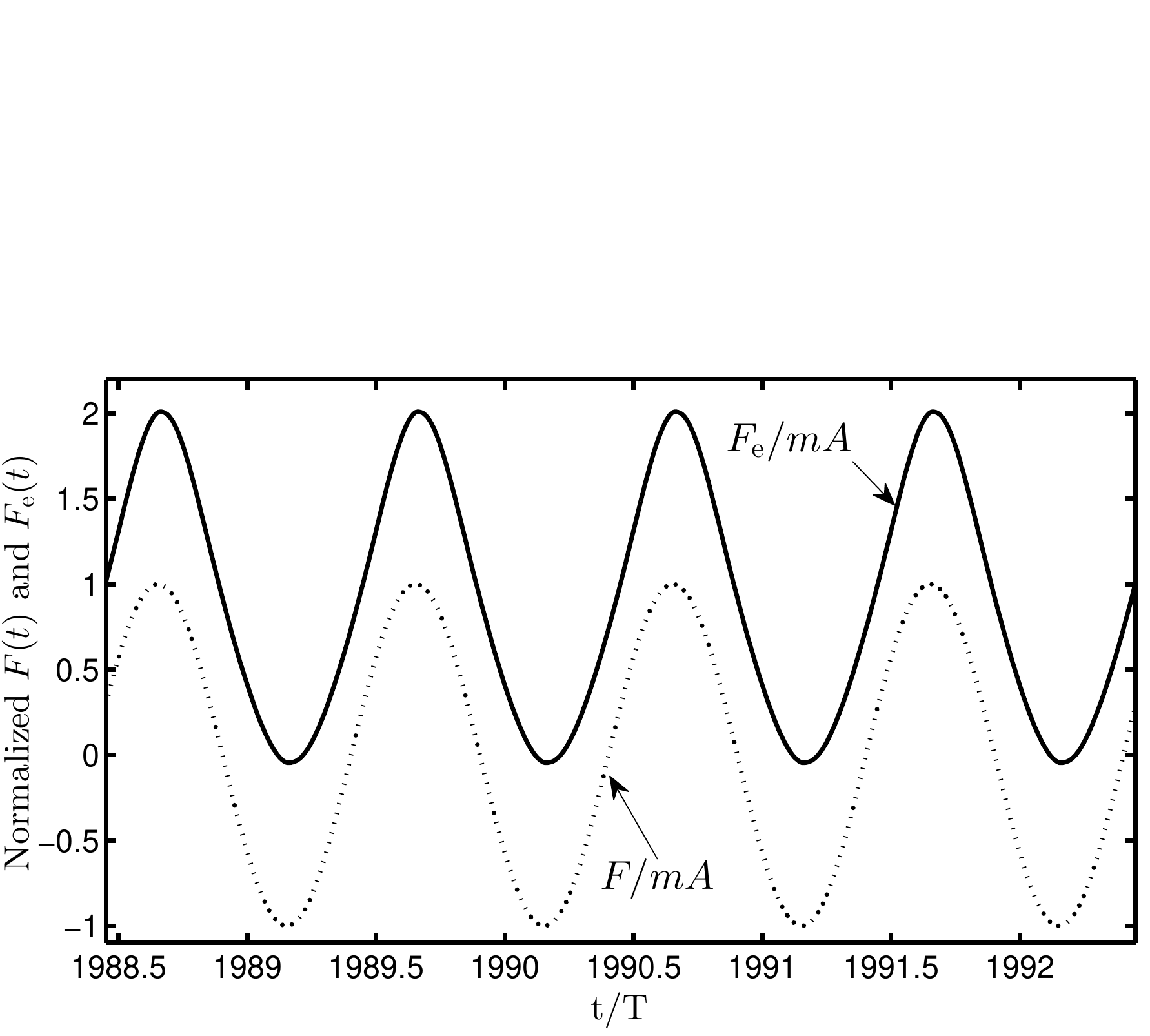}
	\caption{Normalized waveforms of the input acceleration in comparison with the electrostatic force obtained from simulation.}
	\label{Fig:Fin_Fe}
\end{figure}

Figure \ref{Fig:Fin_Fe} shows the comparison between the input acceleration and the electrostatic force, at the same time duration as Figure \ref{Fig:Math_wave}. These simulation results along with expression of $F_\mathrm{e}$ in \eqref{Eq:Fe} confirm that our assumption is reasonable.

The differential equation of the spring-mass-damping system, which is set in continuous oscillation by a sinusoidal force acting on the mass, is
\begin{align}
\small
m \ddot{x} + b  \dot{x} +kx = mA \sin  \big(\omega t \big) - F_\mathrm{e}. \label{Diff_eq}
\end{align}
The steady-state solution of \eqref{Diff_eq} is $x = -\bar{x} + x_\mathrm{h}$, where $\bar{x} = \frac{F_0}{k}$ is the offset displacement and the harmonic term is \cite{Bao2000}
\begin{align}
\small
x_\mathrm{h} =  X_\mathrm{0} \mathrm{sin} \big(\omega t +   \varphi  \big).
\label{x_sol}
\end{align}
where $X_\mathrm{0}= \frac{(mA - F_0)/m}{ \sqrt{  \big(  \omega ^2 - \omega _\mathrm{0}^2\big)^2 +   \big( \frac{b}{m} \big)  ^2   \omega ^2 } } $. Since $\omega = \omega_\mathrm{0} = \sqrt{ \frac{k}{m} }$ and the proof mass displacement barely reaches its constraint, the peak value of $x_\mathrm{h}$ is $X_\mathrm{0} = X_\mathrm{s} = \frac{ mA - F_0}{ b \omega_\mathrm{0}}$. The ratio $\frac{\bar{x}}{X_0}$ obtained from simulations is less than $2.1 \%$ for all $A \in [1,2]$ g, therefore $\bar{x}$ is assumed negligible.
By considering amplitudes of the harmonic term and ignoring phase differences, the saturation voltage is
\begin{align}
\small
V_\mathrm{s} &= \sqrt{ \frac{8}{2 + \sqrt{5}} \frac{mA - \frac{3}{2} \big( \sqrt{5} - 2 \big) x_0 b \omega_\mathrm{0} }{  \frac{C_\mathrm{0}}{x_\mathrm{0}}}}.
\label{Eq:Solution_Ideal}
\end{align}

Although performance of the harvesting system using mathematically ideal diode is analyzed, the power loss due to diode imperfections such as leakage current and junction capacitance is still an open room for investigation. This issue will be explored in the next section.

\section{Operation of the Bennet's Doubler with Schottky diode}

\subsection{Approximated Q-V Cycle at steady state}

\begin{figure}[!tbp]
	\centering
	\includegraphics[width=0.45\textwidth]{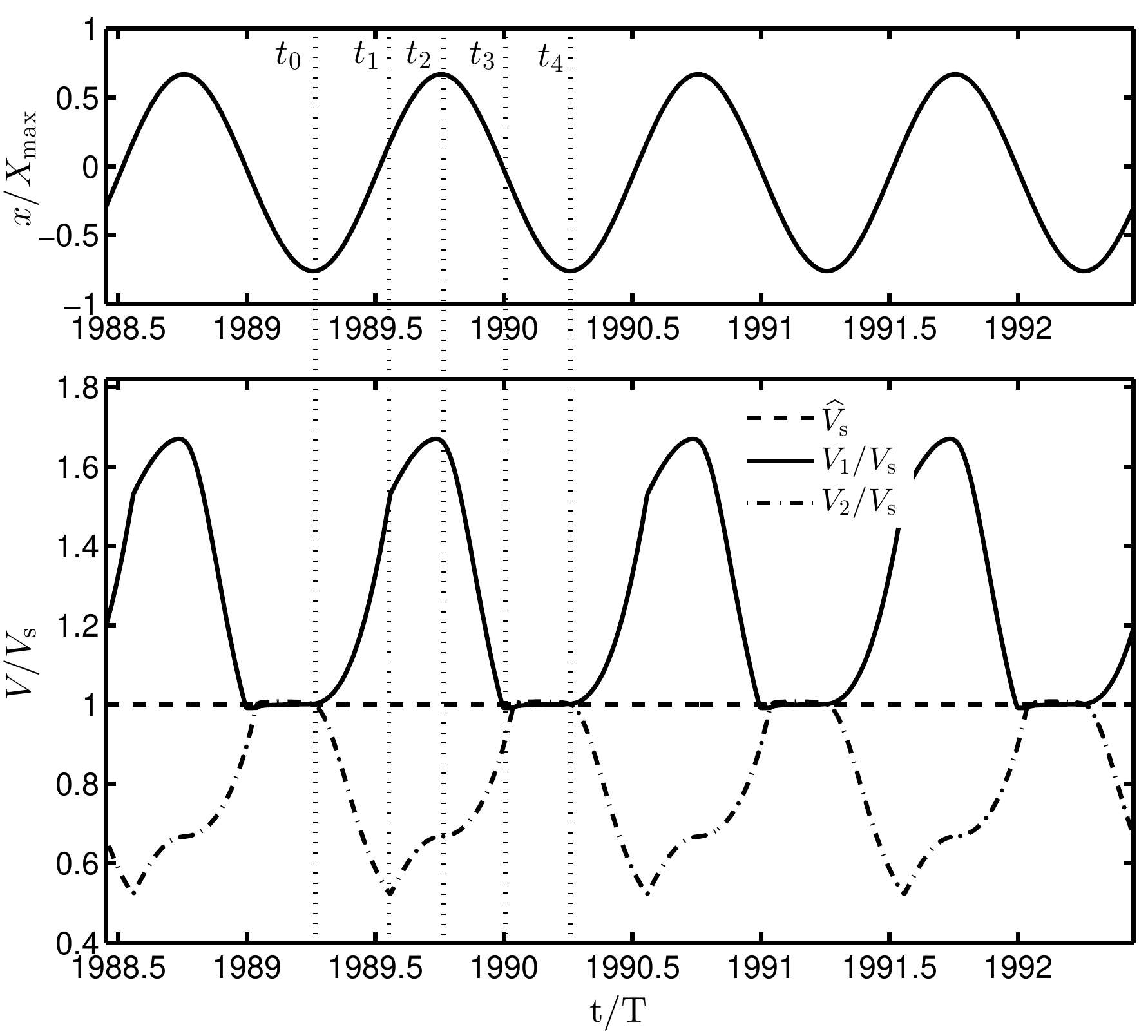}
	\caption{Normalized displacement and voltage waveforms on variable capacitors $C_\mathrm{1} (x)$ and $C_\mathrm{2} (x)$ at saturation with the input external acceleration $A=2.0$ g.}
	\label{Fig:X_V_wave}
\end{figure}

In the same manner of the Dragunov's work \cite{Dragunov2013, Dragunov2015}, the Schottky diode 1N6263 is used to assess effect of diode losses on the harvesting system performance, where the magnitude of reverse current is comparable with the charging current through the storage capacitor, and the zero bias junction capacitance is in the range of transducer nominal capacitance.

%
%
\begin{figure}[!tbp]
	\begin{subfigure}[b]{0.35\textwidth}
		\includegraphics[width=\textwidth]{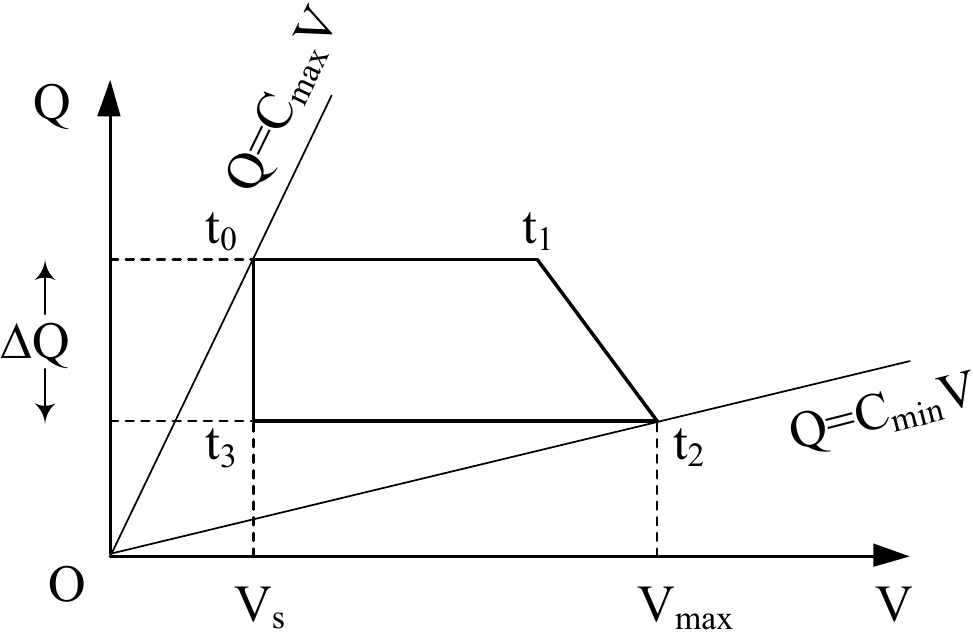}
		\caption{$C_1 (x)$}
		\label{fig:QV1}
	\end{subfigure}
	\hspace{0.02\textwidth}
	\begin{subfigure}[b]{0.35\textwidth}
		\includegraphics[width=\textwidth]{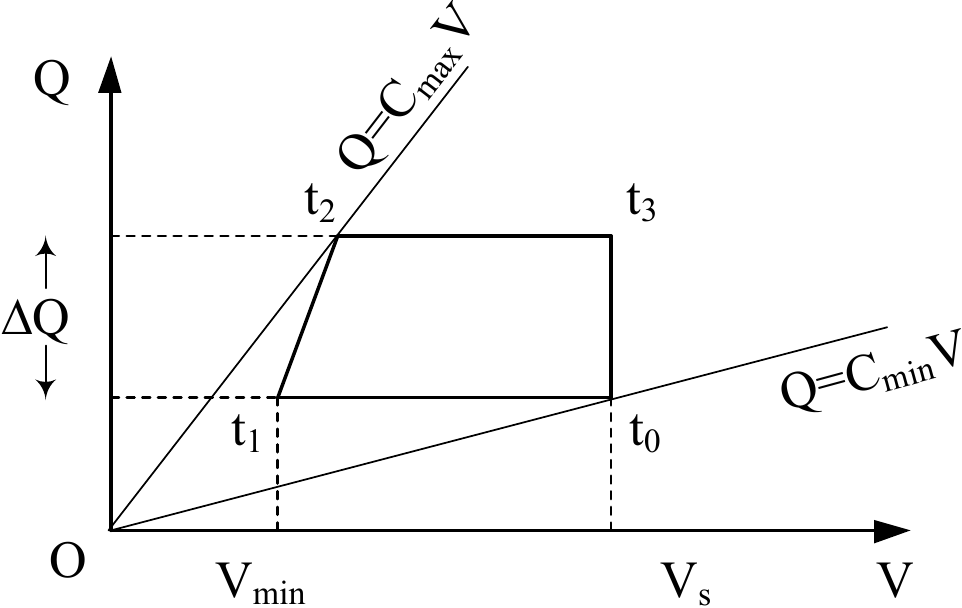}
		\caption{$C_2 (x)$}
		\label{fig:QV2}
	\end{subfigure}
	\caption{Generally approximated Q-V diagram of Bennet's doubler at saturation for both transducers.}
	\label{Fig:QV}
\end{figure}
Figure \ref{Fig:X_V_wave} shows waveforms of the proof mass displacement and the voltages $V_1$ and $V_2$ across $C_\mathrm{1}$ and $C_\mathrm{2}$ respectively. Similarly, operation of the doubler circuit at steady state can be divided into a sequence of four stages, which is more clearly than considerations of mathematically idealized diode (i.e., the time interval between stages is significant). In general, the relation of $Q_1 \, (Q_2)$ and $V_1 \, (V_2)$ at steady state can be approximated by a right-angled trapezoid Q-V cycle diagram in Figure \ref{Fig:QV}. Charges transferred from or into $C_\mathrm{s}$ are neglected since the output voltage is unchanged at steady state. Differently from previous section, the proof mass displacements at $t_1$ and $t_3$ are still unknown.

\textbf{\textit{Stage I}}:

The same as previous analysis, the charges on the two generators and variations of $V_1$ and $V_2$ from $t_0$ to $t_1$ are presented by equations \eqref{q1_t0}, \eqref{q2_t0}, \eqref{V_1} and \eqref{V_2}.

\textbf{\textit{Stage II}}:

At $t=t_1$, $x(t_1) = x_1$, $V_1 (t_1) = V_2 (t_1) + V_\mathrm{s}$ and diode $D_3$ starts to conduct, this yields
\begin{align}
\small
\frac{1 + \frac{C_0}{C_\mathrm{p}}\Big( 1 + \frac{X_\mathrm{s}}{x_0} \Big)}{1 + \frac{C_0}{C_\mathrm{p}} \Big( 1 - \frac{x_1}{x_0} \Big)} = 1 + \frac{1 + \frac{C_0}{C_\mathrm{p}}\Big( 1 - \frac{X_\mathrm{s}}{x_0} \Big)}{1 + \frac{C_0}{C_\mathrm{p}} \Big( 1 + \frac{x_1}{x_0} \Big)}.
	\label{D3}
\end{align}
From $t_\mathrm{1}$ to $t_\mathrm{2}$, charge $\Delta Q$ is pumped from $C_1$ into $C_2$.

\textbf{\textit{Stage III}}:

From $t_2$ to $t_3$, all diodes are blocked, $q_1$ and $q_2$ are constants that are described by \eqref{Ideal:q1_t2_t3} and \eqref{Ideal:q2_t2_t3}.
At $t=t_3$, $D_2$ starts to conduct due to $V_1(t_3) = V_\mathrm{s}~ \refstepcounter{equation}(\theequation) \label{V1_Vs}$. This condition is expressed by \eqref{Ideal:V1_t3},
which results in
\begin{align} \label{Delta_Q}
\small
\Delta Q = V_\mathrm{s} C_0 \Big( \frac{X_\mathrm{s} + x_3}{x_0} \Big).
\end{align}
The voltage across $C_2$ at $t_3$ is
\begin{align}
\small
\begin{split}
V_2 (t_3) = \frac{q_2(t_3)}{C_2 (t_3)} = \frac{ V_\mathrm{s} \Big[C_\mathrm{p} + C_0 (1 - \frac{X_\mathrm{s}}{x_0})\Big] + \Delta Q}{C_\mathrm{p} + C_0 (1 + \frac{x_3}{x_0})} 
= \frac{ V_\mathrm{s} \Big[C_\mathrm{p} + C_0 (1 - \frac{X_\mathrm{s}}{x_0})\Big] + V_\mathrm{s} C_0 \Big( \frac{X_\mathrm{s} + x_3}{x_0} \Big)}{C_\mathrm{p} + C_0 (1 + \frac{x_3}{x_0})} = V_\mathrm{s}.
\end{split}
\end{align}
As the condition $V_2 = V_\mathrm{s}$ is fulfilled, $D_1$ also starts to conduct at $t_3$. Substituting \eqref{Delta_Q} to \eqref{Ideal:q1_t2_t3}, we get
\begin{align}
\small
q_1(t_3) = V_\mathrm{s} \Big[C_\mathrm{p} + C_0 (1 + \frac{x_3}{x_0})\Big].
\end{align}

\textbf{\textit{Stage IV}}:

From $t_3$ to $t_4$, $D_1$ is conducting and $\Delta Q$ is transfered into $C_1$ from $C_2$. At $t_4$, $x (t_4) = -X_\mathrm{s} = x (t_0)$ and the state of the doubler circuit is the same as when $t=t_0$, leading to
\begin{align}
\small
q_1 (t_4) = q_1 (t_0)
\label{q1_t4_t0}
\end{align}
where
\begin{align}
\small
q_1(t_4) = q_1(t_3) + \Delta Q = V_\mathrm{s} \Big[C_\mathrm{p} + C_0 (1 + \frac{X_\mathrm{s} + 2 x_3}{x_0})\Big]. 
\label{q1_t4}
\end{align}
From the equations \eqref{q1_t0}, \eqref{q1_t4_t0} and \eqref{q1_t4}, the displacement at $t_3$ is given by $x_3 = 0$. As the consequence
\begin{align} \label{Delta_Q_Solution}
\small
\Delta Q = V_\mathrm{s} C_0 \frac{X_\mathrm{s}}{x_0}.
\end{align}
Substituting this result back into \eqref{Ideal:q1_t2_t3} and \eqref{Ideal:q2_t2_t3}, the voltages across $C_1$ and $C_2$ at $t_2$ are obtained
\begin{align}
\small
V_1 (t_2) &= \frac{q_1(t_2)}{C_1 (t_2)} = \frac{ V_\mathrm{s} (C_\mathrm{p} + C_0)}{C_\mathrm{p} + C_0 (1 - \frac{X_\mathrm{s}}{x_0})}, \\
V_2 (t_2) &= \frac{q_2(t_2)}{C_2 (t_2)} = \frac{ V_\mathrm{s} (C_\mathrm{p} + C_0)}{C_\mathrm{p} + C_0 (1 + \frac{X_\mathrm{s}}{x_0})}.
\end{align}
At $t_2$, $D_3$ starts to stop conducting since $V_1$ is slightly less than $V_2 + V_\mathrm{s}$. This relation can be approximated as $V_1 \approx V_2 + V_\mathrm{s}$. Similarly as equation \eqref{D3}, we get
\begin{align}
\small
\frac{1 + \frac{C_0}{C_\mathrm{p}}}{1 + \frac{C_0}{C_\mathrm{p}} \Big( 1 - \frac{X_\mathrm{s}}{x_0} \Big)} = 1 + \frac{1 + \frac{C_0}{C_\mathrm{p}}}{1 + \frac{C_0}{C_\mathrm{p}} \Big( 1 + \frac{X_\mathrm{s}}{x_0} \Big)}.
\label{D3_blocked}
\end{align}
The solution of the maximum displacement at steady state is 
\begin{align}
\small
X_\mathrm{s} = \frac{3 (\sqrt{2} - 1)}{2} x_0 .
\label{Xs_sol}
\end{align}
Substituting \eqref{Xs_sol} back into \eqref{D3}, the proof mass displacement at $t_1$ is determined by
\begin{align}
\small
x_1 = \frac{3}{2} \Big(\sqrt{4 - 2\sqrt{2}} - 1\Big) x_0 .
\label{x_t1}
\end{align}
Therefore, the peak values of $V_1$ and $V_2$ are 
\begin{align}
\small
V_\mathrm{I} &= V_1 (t_2) =  V_\mathrm{s} \Big(1 + \frac{1}{\sqrt{2}}\Big),
\label{V1_peak} \\
V_\mathrm{II} &= V_2 (t_1) = V_\mathrm{s} \sqrt{1 - \frac{1}{\sqrt{2}}}. \label{V2_peak}
\end{align}
$V_1$ and $V_2$ are then approximated by
\begin{align}
\small
V_1 &= \frac{V_\mathrm{I} + V_\mathrm{s}}{2} + \frac{V_\mathrm{I} - V_\mathrm{s}}{2} \sin (\omega t) = V_\mathrm{s} (1 + \frac{1}{2 \sqrt{2}}) + V_\mathrm{s} \frac{1}{2 \sqrt{2}} \sin (\omega t),
\label{V1} \\
V_2 &= \frac{V_\mathrm{II} + V_\mathrm{s}}{2} - \frac{V_\mathrm{s} - V_\mathrm{II}}{2} \sin (\omega t) = V_\mathrm{s} \frac{1+ \sqrt{1 - \frac{1}{\sqrt{2}}}}{2} - V_\mathrm{s} \frac{1 - \sqrt{1 - \frac{1}{\sqrt{2}}}}{2} \sin (\omega t)
\label{V2}
\end{align}
yielding
\begin{align}
\small
V_1^2 - V_2^2 & = V_\mathrm{s}^2 \big(\alpha + \gamma \big) \big[ \big(\alpha - \gamma \big) + \big(\beta + \lambda \big) \sin \big( \omega t \big) \big] \Big( 1 + \frac{\beta - \lambda}{\alpha + \gamma} \sin \big( \omega t \big) \Big)
\end{align}
where
\begin{equation}
\small
\alpha = 1 + \frac{1}{2 \sqrt{2}},\, \beta = \frac{1}{2 \sqrt{2}},\, \gamma = \frac{1 + \sqrt{1 - \frac{1}{\sqrt{2}}}}{2}, \, \lambda = \frac{1 - \sqrt{1 - \frac{1}{\sqrt{2}}}}{2}.
\end{equation}
Since $\frac{\beta - \lambda}{\alpha + \gamma} \approx 0.058 \ll 1$ is negligible and $\alpha - \gamma = \beta + \lambda $, the electrostatic force can be given by
\begin{align}
\small
F_\mathrm{e} = \frac{1}{2} \frac{C_0}{x_0} V_\mathrm{s}^2 \big(\alpha^2 - \gamma^2 \big) \big( 1 + \sin (\omega t)\big) &= \frac{1}{2} \frac{C_0}{x_0} V_\mathrm{s}^2 \frac{5 (1+\sqrt{2}) - 2 \sqrt{4 - 2\sqrt{2}}}{8} \big( 1 + \sin (\omega t)\big)
\end{align}
which is represented as
\begin{align}
\small
F_\mathrm{e} = F_0 \big( 1 + \sin (\omega t)\big)
\end{align}
where $F_0 = \frac{5 (1+\sqrt{2}) - 2 \sqrt{4 - 2\sqrt{2}}}{16} \frac{C_0}{x_0} V_\mathrm{s}^2$. 

Using the same analysis procedure in the previous section, the saturation voltage is
\begin{align}
\small
\begin{split}
V_\mathrm{s} = \sqrt{ \frac{16}{5 (1+\sqrt{2}) - 2 \sqrt{4 - 2\sqrt{2}}} \frac{mA - \frac{3}{2} \big( \sqrt{2} - 1 \big) x_0 b \omega_\mathrm{0} }{  \frac{C_\mathrm{0}}{x_\mathrm{0}}}} \approx \sqrt{ 1.61 \frac{mA - \frac{3}{2} \big( \sqrt{2} - 1 \big) x_0 b \omega_\mathrm{0} }{  \frac{C_\mathrm{0}}{x_\mathrm{0}}}} .
\end{split}
\label{Eq:Solution_NonIdeal}
\end{align}

\begin{figure}[!tbp]
	\centering
	\includegraphics[width=0.35\textwidth]{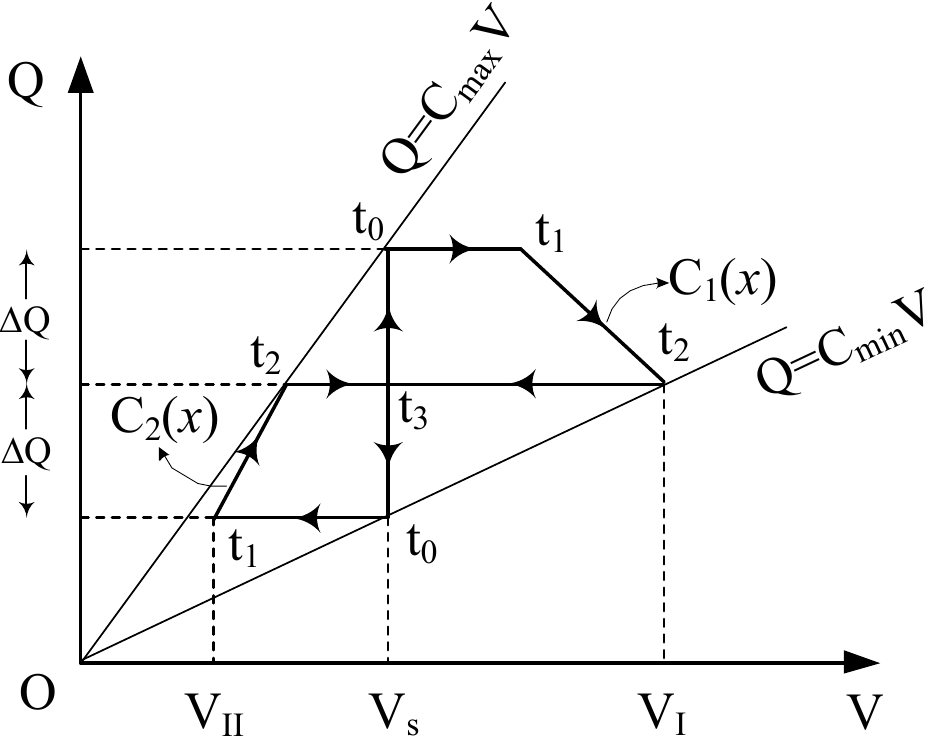}
	\caption{Q-V diagram of Bennet's doubler at steady-state for both variable capacitors.}
	\label{Fig:QV_Full}
\end{figure}
Based on those analysis above, the completed Q-V diagram combined by both transducers is summarized in Figure \ref{Fig:QV_Full}.

\subsection{Numerical validations}

\begin{figure}[!tbp]
	\begin{subfigure}[b]{0.45\textwidth}
		\includegraphics[width=\textwidth]{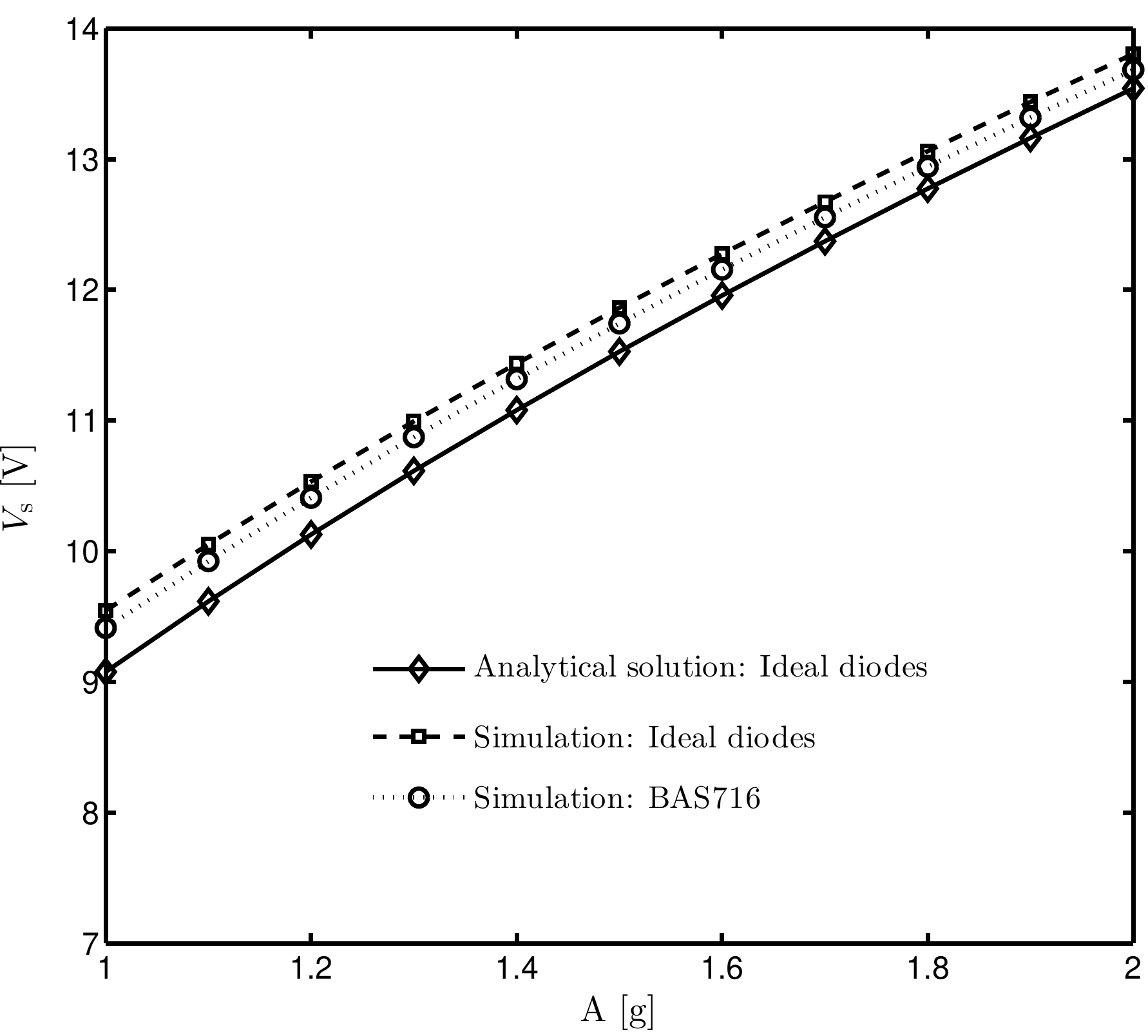}
		\caption{Ideal/Low-losses diode}
		\label{Fig:Vs_Ideal}
	\end{subfigure}
	\hspace{0.02\textwidth}
	\begin{subfigure}[b]{0.45\textwidth}
		\includegraphics[width=\textwidth]{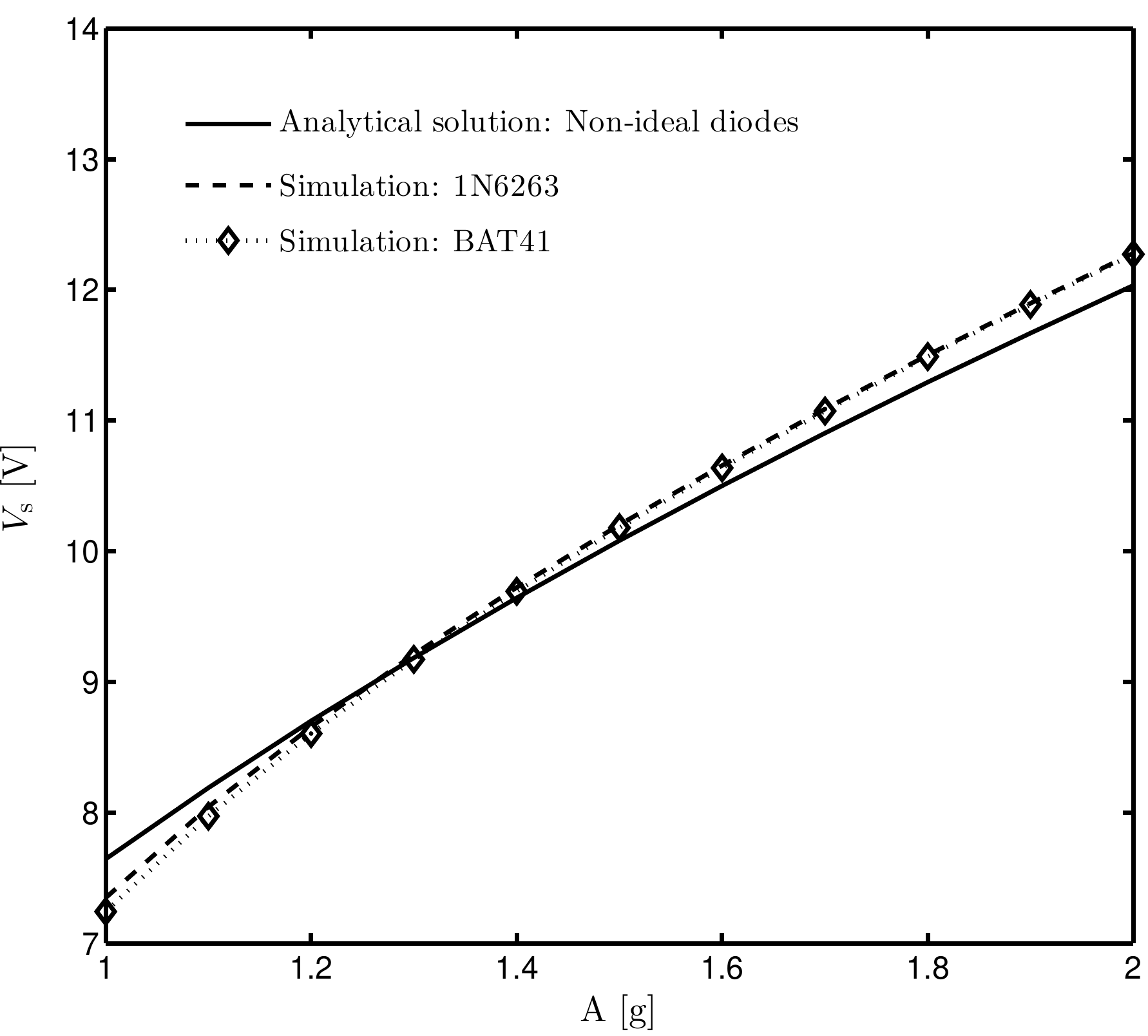}
		\caption{Schottky diode}
		\label{Fig:Vs_NonIdeal}
	\end{subfigure}
	\caption{Acceleration responses of output voltages at steady state: a comparison between simulations and analytical solutions.}
	\label{Fig:Vs}
\end{figure}

\begin{table}[!h]
	\Small
	\centering
	\caption{Diodes parameters: reverse saturation current $I_\mathrm{s}$, zero-bias junction capacitance $C_\mathrm{j}$ and built-in junction voltage $V_\mathrm{j}$}%
	\begin{tabular}{l l l l} 
		\hline\hline
		\textbf{Diode} & \textbf{$I_\mathrm{s}$ [nA]} & \textbf{$C_\mathrm{j}$} [pF] & \textbf{$V_\mathrm{j}$} [V] \\
		\hline
		1N6263 & 3.87 & 1.77 & 0.39\\ 
		BAS716 & 3.52e-6 & 1.82 & 0.65\\
		BAT41 & 10.00 & 5.76 & 0.37\\
		\hline\hline
	\end{tabular}
	\label{Tab:diode} 
\end{table}

Figure \ref{Fig:Vs_Ideal} shows the saturation voltages for different acceleration amplitudes, where the simulation results with use of the mathematically idealized diode and the analytical solution expressed by formula \eqref{Eq:Solution_Ideal} are compared. 
The figure also exhibits that the low-losses diode BAS716 performs very close to that of mathematically idealized diode. 
In the same manner, Figure \ref{Fig:Vs_NonIdeal} presents the comparison of the analytical solution obtained from \eqref{Eq:Solution_NonIdeal} against the numerical simulations using different Schottky diodes.
Despite of disparities in reverse current, junction capacitance and built-in junction voltage, both diodes 1N6263 and BAT41 give almost the same saturation voltages.
The agreement between theoretical and numerical results in both cases verifies the predictions of our analytical approach and solutions.
Diode parameters used on the simulations are listed in Table \ref{Tab:diode}.

\subsection{Effect of diode operation on mechanical dynamics}

The Q-V cycle is a useful geometrical tool that enables us to realize the operation of voltage doubler circuit at steady state. However, the harvesting system performance in reality is more sophisticated, especially in transient time.


Based on dynamic simulations, we observe that the phases of the external force $F(t)$, the proof mass displacement $x(t)$ and the electrostatic force $F_\mathrm{e}(t)$ are initially different. However, those differences gradually decrease due to effect of the diode states (i.e. blocked and conducting). This variation process leads to the negligible phase shift at steady state. Such a clarification supports the assumption that we made in theoretical analysis sections. In other words, the dynamic motion of the proof mass also strongly depends on both of the transducing force and the diode operation mechanism. This statement is valid when different diode models such as the mathematically idealized diode and the Schottky diodes are utilized.


\section{Circuit topologies to improve the saturation voltage}

\subsection{A new voltage doubler with single switch } \label{SingleSW}

\begin{figure}[!tbp]
	\centering
	\includegraphics[width=0.3\textwidth]{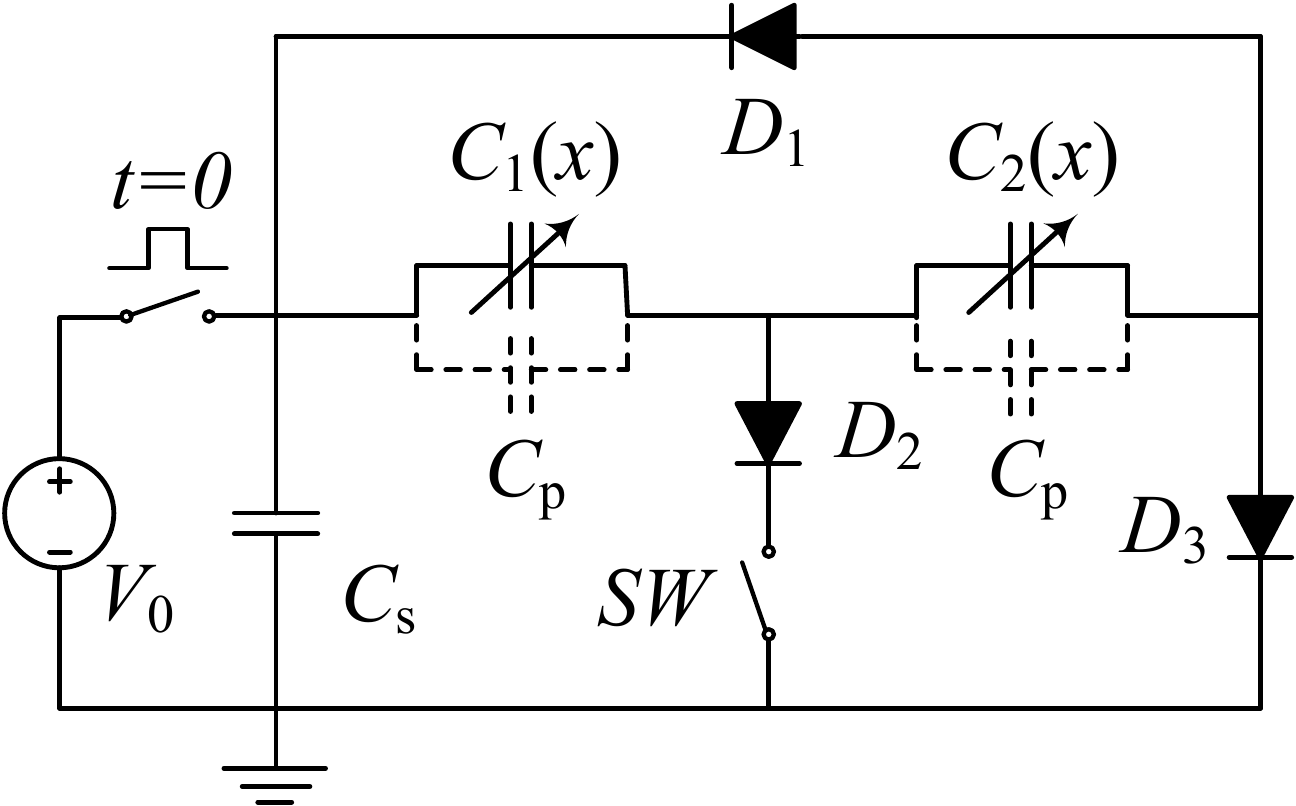}
	\caption{A new topology with single switch connected in series with the diode $D_2$.}
	\label{Fig:with_SW}
\end{figure}

\begin{figure}[!tbp]
	\begin{subfigure}[b]{0.45\textwidth}
		\includegraphics[width=\textwidth]{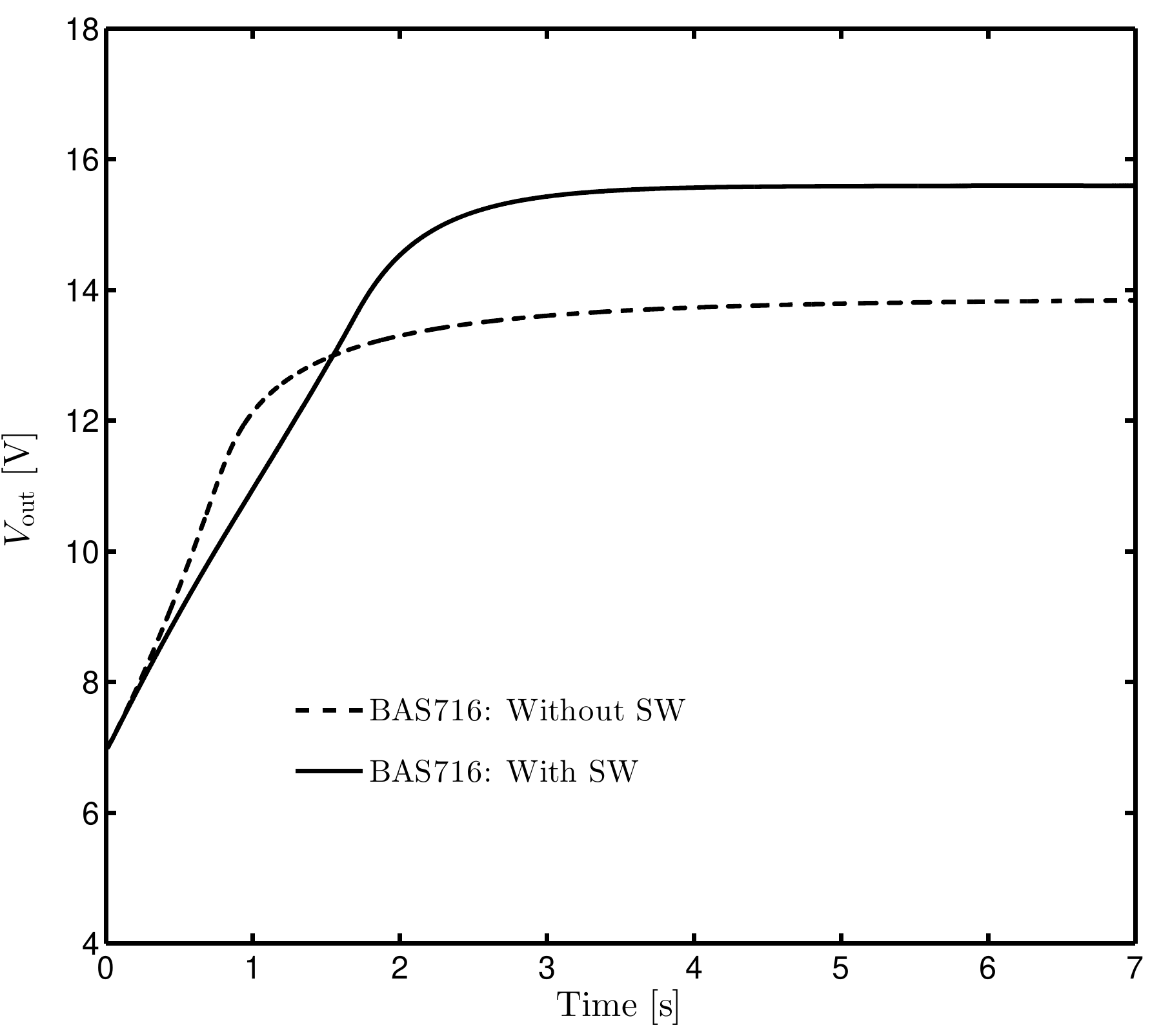}
		\caption{$A=2$ g}
		\label{Fig:With_Without_SWa}
	\end{subfigure}
	\hspace{0.02\textwidth}
	\begin{subfigure}[b]{0.45\textwidth}
		\includegraphics[width=\textwidth]{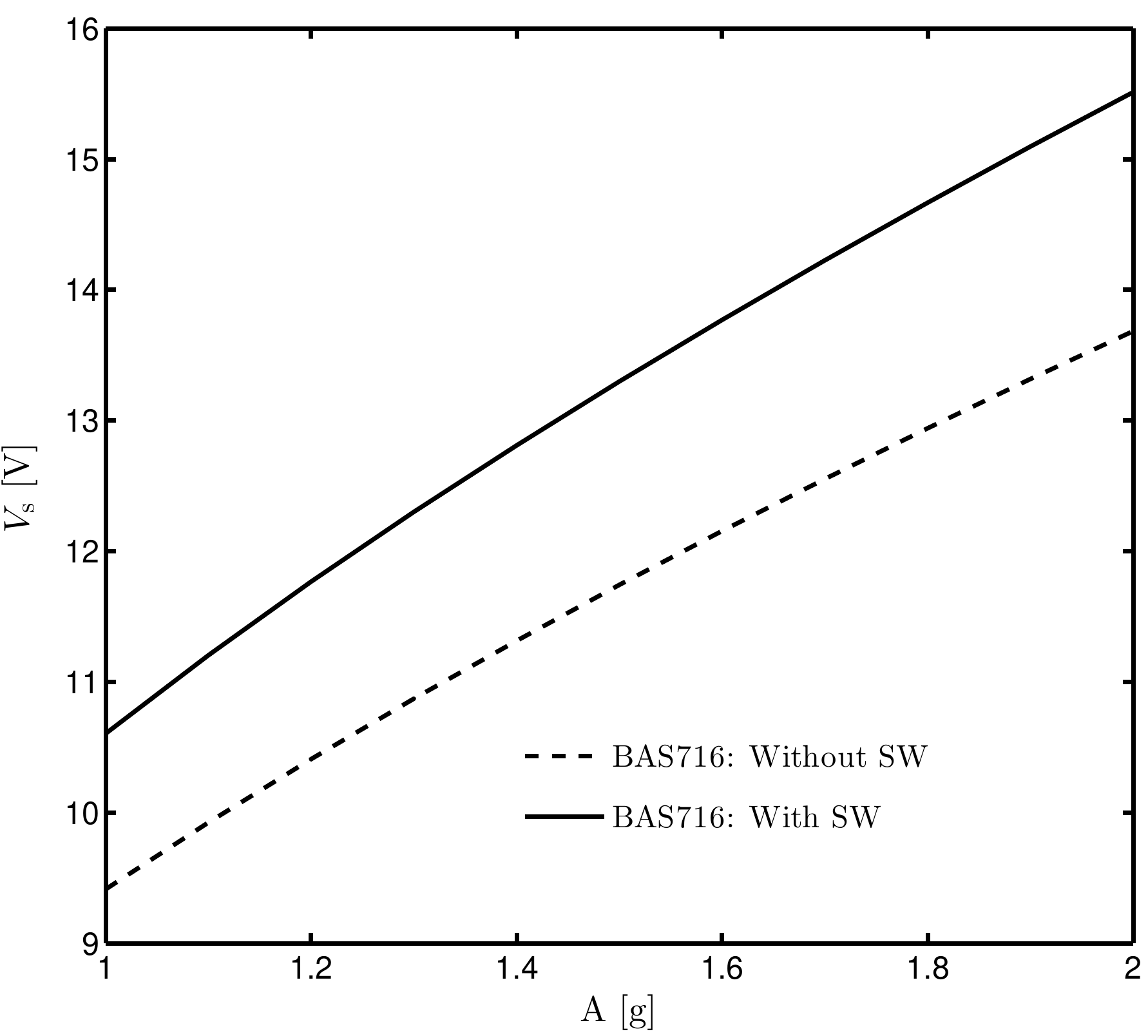}
		\caption{Acceleration responses}
		\label{Fig:With_Without_SWb}
	\end{subfigure}
	\caption{(a) The time evolution of output voltage at $A=2$ g and (b) The saturation voltage versus acceleration amplitudes, comparison of two cases: with and without the switch.}
	\label{Fig:With_Without_SW}
\end{figure}

Although the diode $D_2$ plays an vital role for initially charging $C_1$, in principle, it could be removed after a few transient vibration cycles. This also enlarges the charging current through the storage capacitor due to the relation $I_\mathrm{Cs} = I_\mathrm{D3} - I_\mathrm{D2}$. 
Therefore, it is worthwhile to investigate performance of the harvester when $D_2$ is disconnected. An electronic switch $SW$ in series with $D_2$ can be used for this function, as shown in Figure \ref{Fig:with_SW}.

In the simulation, $SW$ is only \textit{ON} in the first several vibration cycles, then turned \textit{OFF} to eliminate effect of $D_2$ on $I_\mathrm{Cs}$. Figure \ref{Fig:With_Without_SWa} shows evolution of the output voltage in two cases without and with presence of $SW$. Saturation voltage in the latter case is about $\sim 15.60$ V. This is a significant improvement over the 13.84 V achieved for the circuit topology in Figure \ref{Fig:Lumped-model}. Similar results are obtained with different acceleration amplitudes in Figure \ref{Fig:With_Without_SWb}.

\subsection{Cockcroft-Walton generator applied to MEMS device}

\begin{figure}[!tbp]
	\centering
	\includegraphics[width=0.32\textwidth]{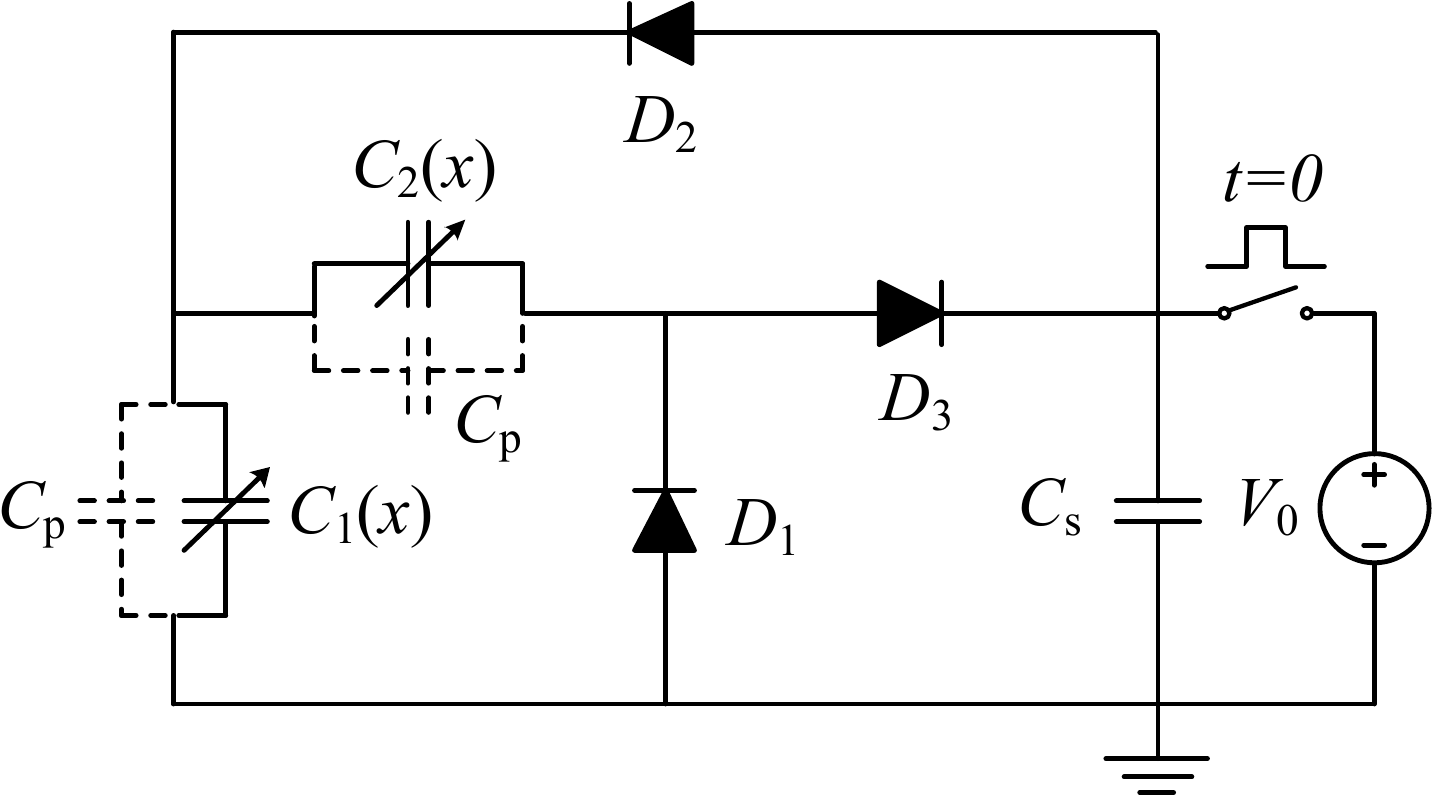}
	\caption{An adapted configuration of the Greinacher's doubler.}
	\label{Fig:Greinacher}
\end{figure}
A common topology of voltage doubler further developed from the Greinacher circuit \cite{Hauschild2014} is depicted in Figure \ref{Fig:Greinacher}, in which the feed-back diode $D_2$ is added to connect the storage capacitor and the two transducers. Both theoretical operation analysis and simulation results show that performances of the Bennet's doubler and the Greinacher configuration are completely identical. The roles of three diode $D_1,\, D_2$ and $D_3$ are the same as they do in Figure \ref{Fig:Lumped-model}.

\begin{figure}[!tbp]
	\centering
	\includegraphics[width=0.5\textwidth]{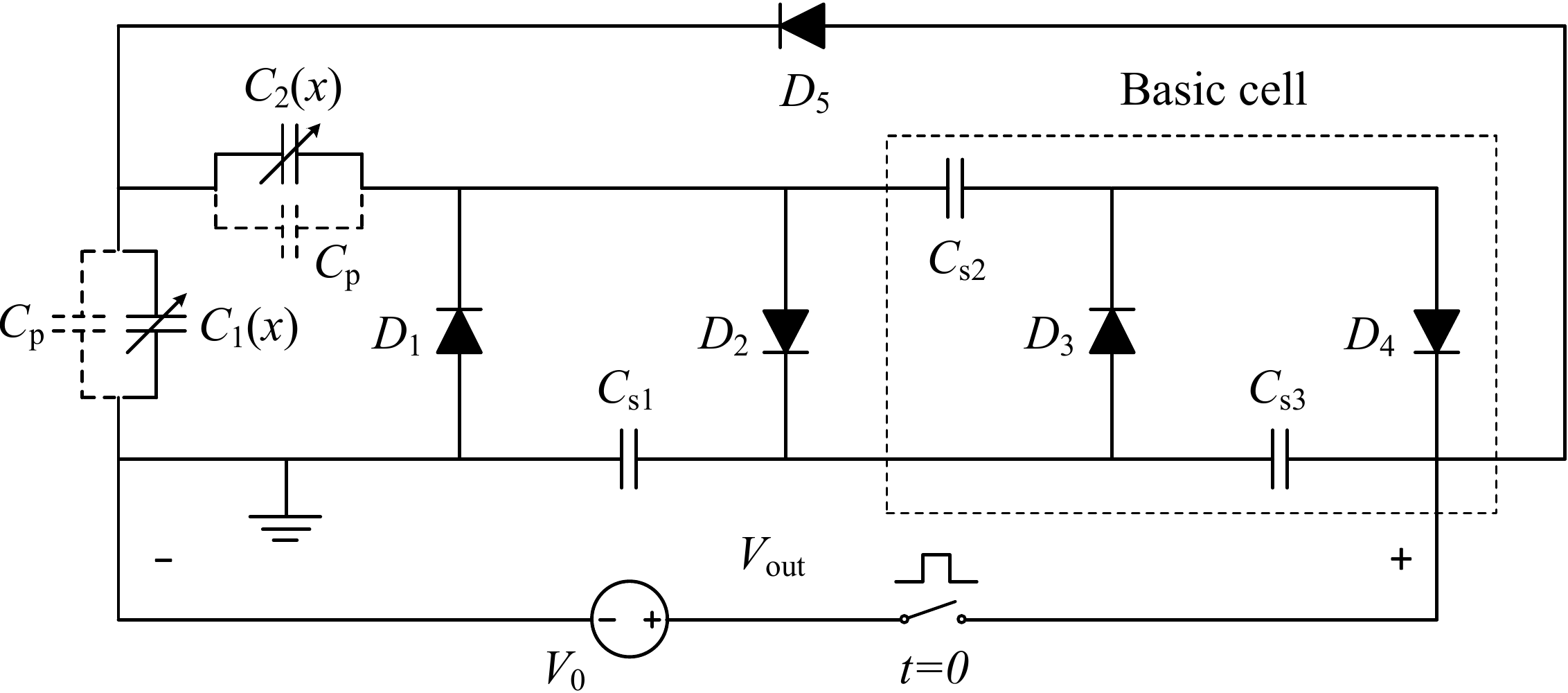}
	\caption{The two-stage Cockcroft-Walton multiplier.}
	\label{Fig:Cockcroft-Walton}
\end{figure}
\begin{figure}[!tbp]
	\centering
	\includegraphics[width=0.5\textwidth]{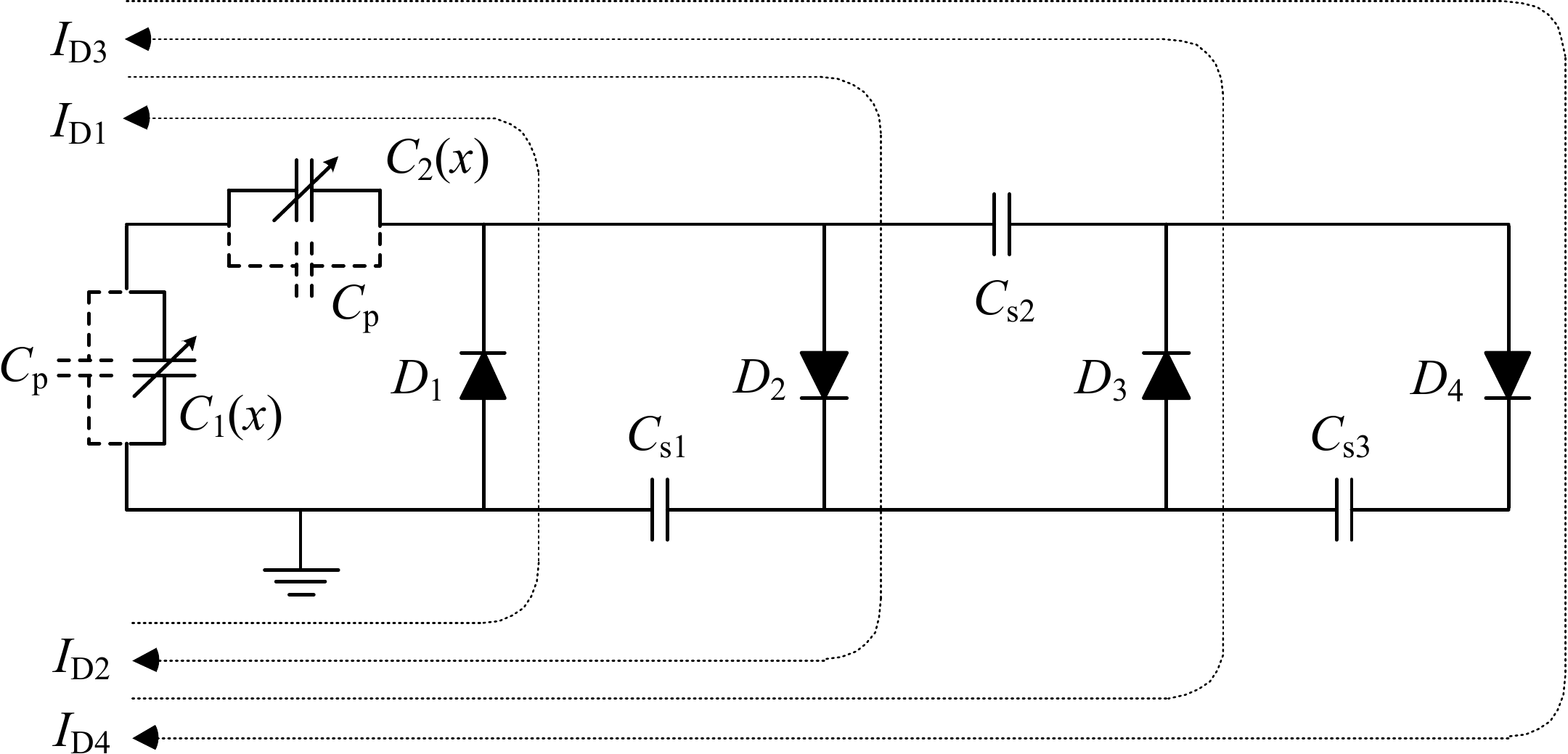}
	\caption{Main operation of the two-stage Cockcroft-Walton multiplier.}
	\label{Fig:CW_Operation}
\end{figure}
Based on the Greinacher doubler circuit, a well-known voltage cascade was early proposed by the British and Irish physicists John D. Cockcroft and Ernest T. S. Walton in 1932 \cite{Cockcrofta, Cockcroftb}. The Cockcroft-Walton generator (i.e., named after the two authors) was proved to be able to generate a high DC voltage from a low-voltage AC, which therefore is interesting to be utilized for the micro-scale harvesters. Figure \ref{Fig:Cockcroft-Walton} shows the circuit diagram of the two-stage Cockcroft-Walton, in which the voltage across two capacitor $C_\mathrm{s1}$ and $C_\mathrm{s3}$ is the output voltage, called $V_\mathrm{out}$. The simplified operation of such a multi-stage voltage doubler is depicted in Figure \ref{Fig:CW_Operation}. Similar to the Bennet's configuration, operation of the Cockcroft-Walton multiplier can also divided into a sequence of four stages. At first, all diodes are blocked. In the second stage, $D_1$ and $D_3$ are simultaneously conducting and charges are transferred to $C_2$ and $C_\mathrm{s2}$. All diodes are reverse-biased in the third stage. In the final stage,  $D_2$ and $D_4$ are conducting, transferring the scavenged energy to $C_\mathrm{s1}$ and $C_\mathrm{s3}$. $D_5$ is mainly used for pre-charging $C_1$ and its conduction during operation is insignificant and negligible.

\begin{figure}[!tbp]
	\centering
	\includegraphics[width=0.45\textwidth]{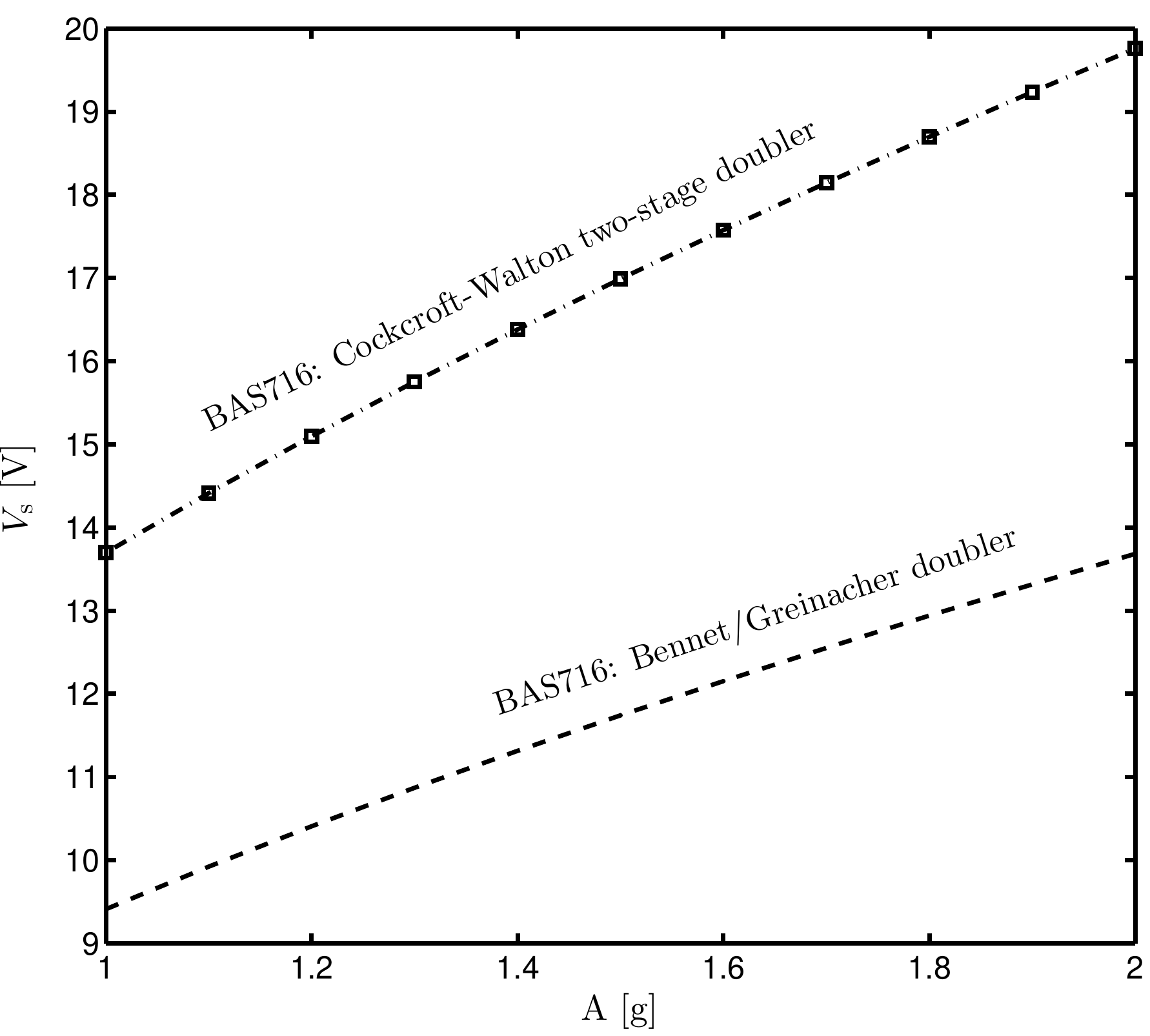}
	\caption{Comparison of the saturation voltage versus acceleration amplitude for the Bennet's doubler and the two-stage Cockcroft-Walton voltage multiplier.}
	\label{Fig:Cockcroft-Walton_Vs}
\end{figure}
Figure \ref{Fig:Cockcroft-Walton_Vs} shows a remarkable increase of the saturation voltage when the Cockcroft-Walton multiplier and the Bennet's doubler are compared. Since the topology discussed in Section \ref{SingleSW} requires a control unit for controlling the switch, the Cockcroft-Walton multiplier is much more convenient to keep the simplicity in practical implementation. Furthermore, our simulations reveal that this circuit topology is capable of operating with very low ratio of capacitance variation $\eta < 2$. In particular, its minimum value is found $\eta_\mathrm{min}=1.52$, making such a circuit attractive for further investigation in future work.

\section{Conclusion}

This study presented a theoretical analysis of MEMS electrostatic energy harvesters configured as Bennet's doubler at saturation regime, based on combination of Q-V diagram and dynamic simulations. 
The steady state operation of voltage doubler was approximately determined as a right-angled trapezoidal conversion cycle.
Mathematically idealized and non-ideal diode models were investigated, resulting in different analytical solutions of the saturation voltages.
The theoretical approach was verified by circuit simulation results obtained from a complete model of the harvesting system.
An essential effect of the diode operation mechanism to the in-phase behavior of the input mechanical vibration and the electrostatic force was discussed. A similarity of Bennet's doubler and resistive fly-back charge-pump circuit is realized by comparing their Q-V diagram. An alternative circuit using a single switch was introduced, where the saturation voltage was significantly improved in comparison with the conventional topologies. The Cockcroft-Walton multiplier is another promising solution since it shows a potential to work with MEMS harvesters that have small varying capacitance ratio.

\vfill

\end{document}